\documentclass[%
 reprint,
 amsmath,amssymb,
 aps,prl,superscriptaddress,
floatfix,
]{revtex4-1}

\usepackage{graphicx}
\usepackage{dcolumn}
\usepackage{bm}
\usepackage{hyperref}
\usepackage{xcolor}
\usepackage{soul}

\begin{document}

\title{Non-monotonous translocation time of polymers across pores}

\author{Emanuele Locatelli}
\email{emanuele.locatelli@unipd.it}
\affiliation{Dipartimento di Fisica e Astronomia, Università di Padova, via Marzolo 8, I-35131 Padova, Italy}
\affiliation{INFN, Sezione di Padova, via Marzolo 8, I-35131 Padova, Italy }%
\author{Valentino Bianco}
\affiliation{Faculty of Chemistry, Chemical Physics Department, Universidad Complutense de Madrid,28040 Madrid, Spain}
\author{Chantal Valeriani}
\affiliation{Departamento de Estructura de la Materia, Física Termica y Electronica, Facultad de Ciencias Físicas, Universidad Complutense de Madrid, 28040 Madrid, Spain}
\author{Paolo Malgaretti}
\email{p.malgaretti@fz-juelich.de}
\affiliation{Helmholtz Institut Erlangen-N\"urnberg  for Renewable Energy (IEK-11), Forschungszentrum J\"ulich, Cauer Str. 1, 91058, Erlangen, Germany}

\date{\today}

\begin{abstract}
Polymers confined in corrugated channels, i.e. channels of varying amplitude, display {multiple local maxima and minima 
of the diffusion coefficient upon increasing their degree of polymerization $N$}. We propose a theoretical effective free energy for linear polymers based on a Fick-Jacobs approach. We validate the predictions against numerical data, obtaining quantitative agreement for 
{the effective free energy, the diffusion coefficient and the Mean First Passage Time}. Finally, we employ the effective free energy to compute the polymer lengths $N_{min}$ at which the diffusion coefficient presents a minimum: we find a scaling expression that we rationalize with a blob model. Our results could be useful to design porous adsorbers, that separate polymers of different sizes without the action of an external flow.      
\end{abstract}

\maketitle

The transport of polymers across corrugated channels and pores is of capital importance for several biological scenarios and technological applications, such as cell regulation~\cite{Albers_book}, DNA and protein sequencing~\cite{Reisner2012,Rathnayaka2022,Floyd2022} and polymer separation~\cite{Sonker2019}.  
Further, polymer transport across corrugated channels is still an open challenge~\cite{Ding2021} since  it couples  all polymer's degrees of freedom (from the monomer  up to the full chain length) to the channel geometry in a non-trivial manner. 
Such an interplay can lead to surprising phenomena. Polymers confined within nanoscopic cylindrical pores exhibit an enhanced mobility as compared to polymers in  bulk~\cite{Shin2007} and nano-channel translocation of DNA can be enhanced by tailoring three-dimensional nano-funnels at the channel entrance~\cite{Zhou2017}; similar entropic traps have been used to purify~\cite{Agrawal2018} and separate~\cite{Han1999,Liu1999,Han2000,Lee2007} DNA and to induce giant enhancement in polymer diffusion~\cite{Kim2017}. \\
Despite its interest, an overall understanding of polymer transport across  channels of varying section is still lacking. 
On the theoretical side, while much attention has been paid to the case of polymer translocation across a pinhole of the size of the monomers~\cite{Muthukumar_book} the general problem of translocation across varying-section pores has not yet been fully addressed. The difficulty relies on the fact that, compared to the former case in which the polymer has to cross "head-first", in the latter case the polymer can cross the  pore's  bottleneck in a variety of configurations. The translocation  probability will depend on monomer-monomer as well as on the monomer-walls effective interactions. 
In this regime, some analytical results have been derived for  "short" polymers, whose gyration radius is much smaller than the distance $L$ between subsequent bottlenecks~\cite{bianco2016}; numerical results have been derived for both linear~\cite{Mutukumar2016} and ring polymers~\cite{marenda2017sorting}.\\
\begin{figure}[t]
\centering
 \includegraphics[width=0.35\textwidth]{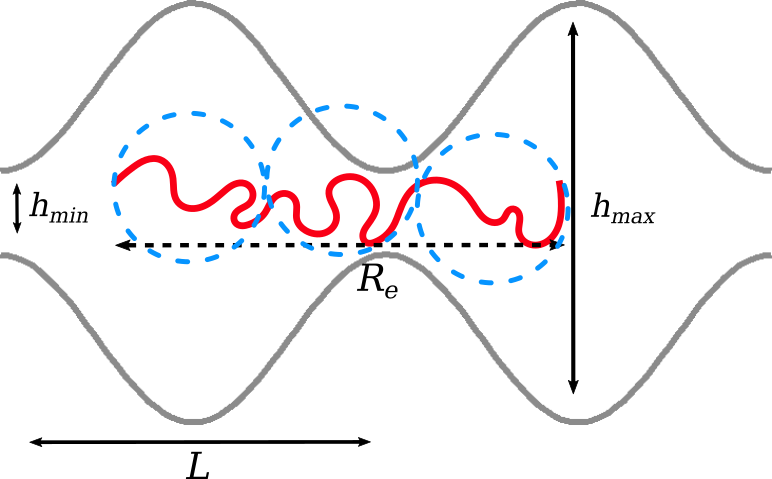}
\caption{Sketch of a linear polymer confined in a corrugated channel, whose radius changes from $h_{\rm min}$ at the bottleneck to $h_{\rm max}$ in the widest section. Dashed circles mark correlated blobs of size $h_0 = (h_{\rm min}$+$h_{\rm max})/2$.}
 \label{fig:model}
\end{figure}
In this Letter, we show that the translocation time {as well as the effective diffusion coefficient} of a linear polymers across a varying-section channel (see Fig.\ref{fig:model}) attains a non-monotonous dependence on the polymer size, $N$.
Remarkably, the deviation of the diffusion coefficient from the expected $1/N$ Rouse dependence can be $10$ to $100$-fold. In order to understand such a behavior we  extend the Fick-Jacobs approximation for (short) polymers~\cite{bianco2016} to the case of arbitrarily long polymers by accounting, in an effective way, for the extension of the polymer inside the corrugated channel. 
Once validated against the numerical data, we exploit our model to predict the polymer size $N_{min}$ at which the diffusion minima occur and we investigate the scaling properties of such quantities with respect to the geometry of the corrugated channel. These features can be captured by a simple blob model, that yields the scaling of $N_{min}$ as a function of the geometry of the channel. Such a classic approach may provide a useful tool to design channels for polymer sorting.

\begin{figure}[t]
  \includegraphics[width=0.4\textwidth]{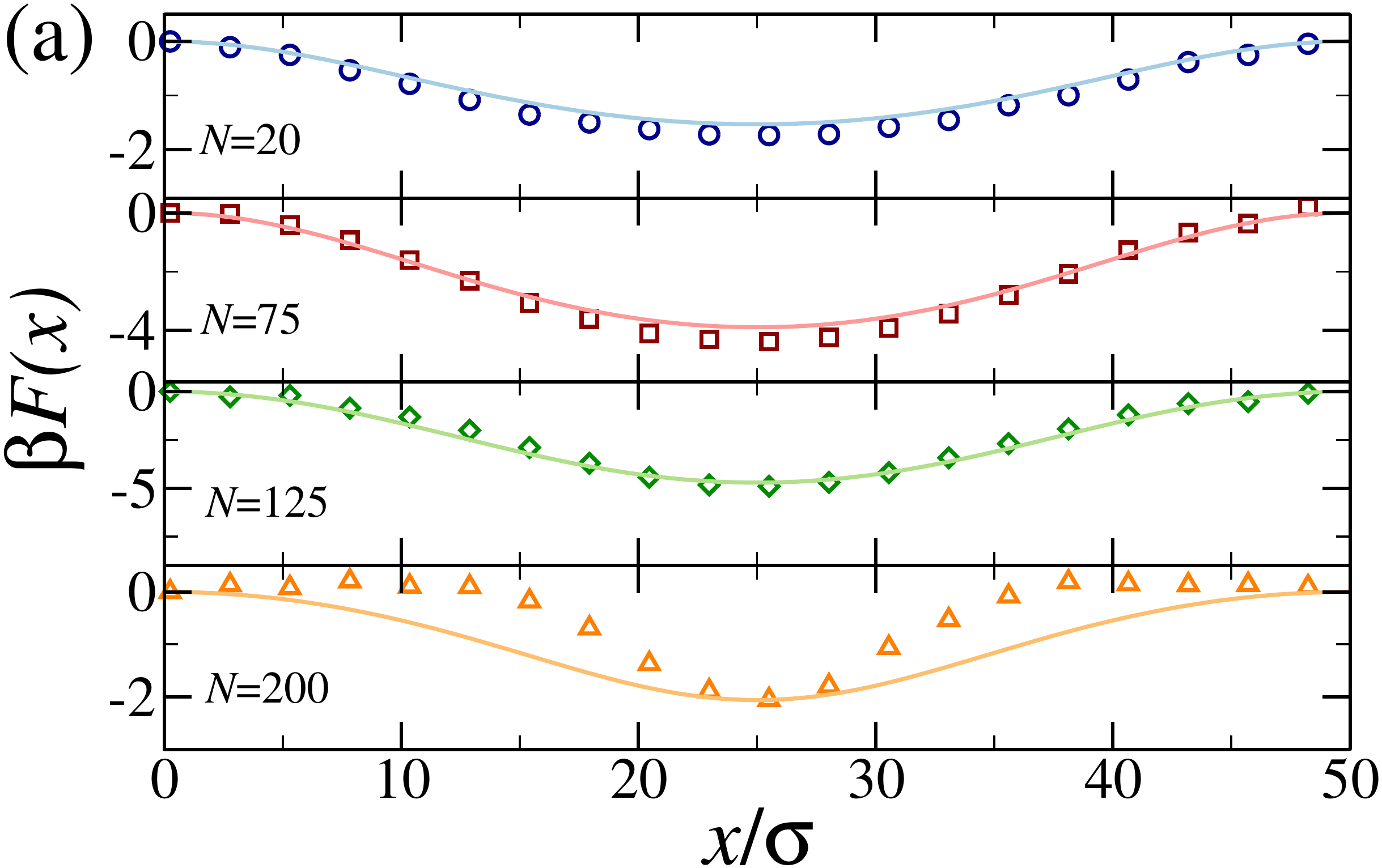}
  \includegraphics[width=0.4\textwidth]{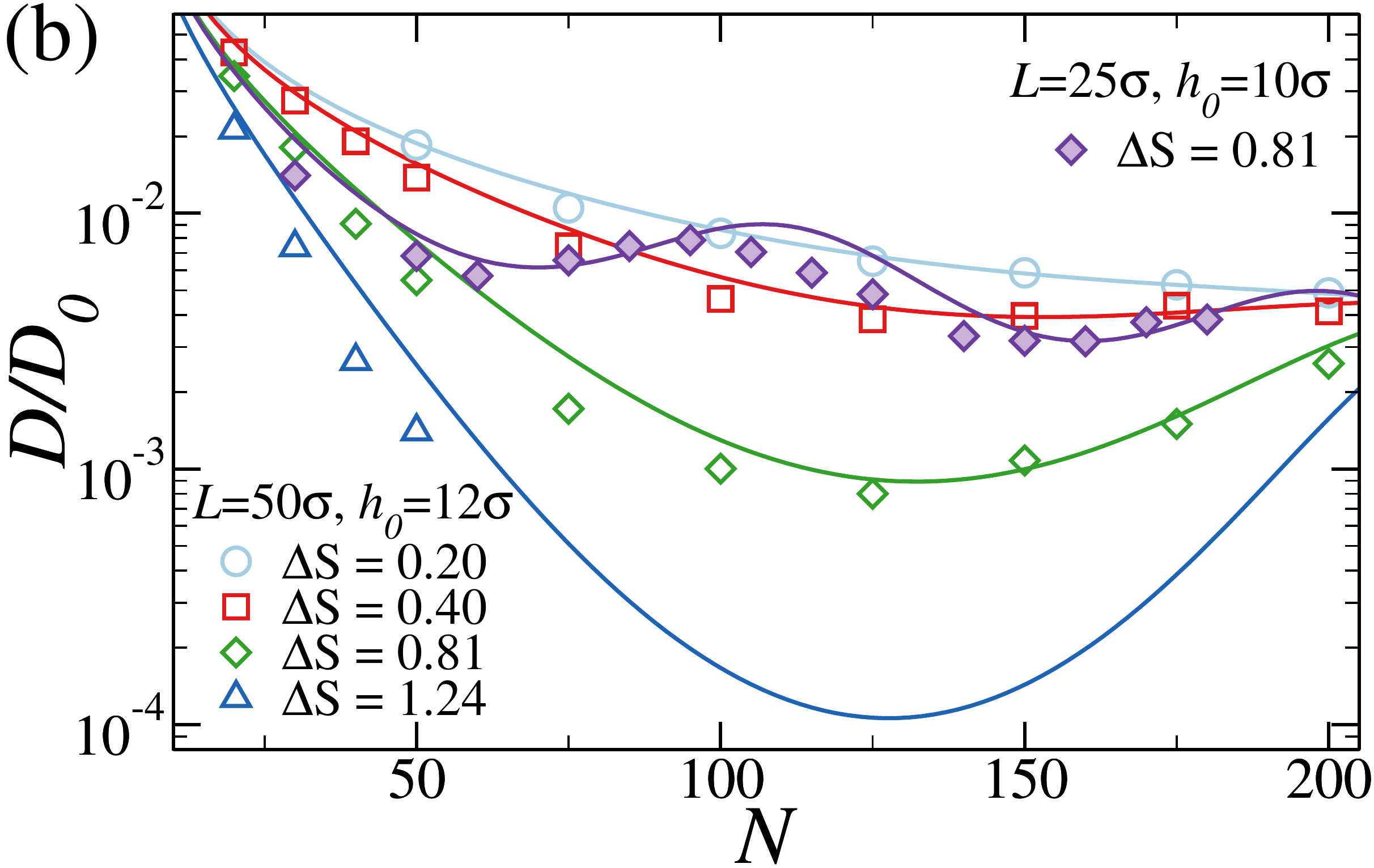}\\
   \caption{      a) Comparison between theoretical and numerical free energy along the channel axis for $L=50 \sigma$, $h_0=12 \sigma$ and $\Delta S= 0.81$. Symbols refer to numerical free energies, lines to Eq.~(\ref{eq:newf}). We shift the curves so that the free energy at the bottleneck is always zero.   
   b) Long time diffusion coefficient $D$, normalized by the diffusion coefficient of a single monomer $D_0$, as function of $N$ for different values of $L$, $h_0$ and $\Delta S$. Symbols refer to numerical data, lines to the theoretical estimation Eqs.~(\ref{eq:newf}), ~(\ref{eq:D_fj}).   
\\ 
   }
\label{fig:free_energy_comparison}
\end{figure}

We perform standard Langevin Dynamics simulations (see Suppl. Mat. 2A)\footnote{See Supplemental Materials at [URL will be inserted by publisher] for additional definitions and theoretical details, model and simulation details, and additional numerical data, including Refs.[34-36].} for channels of length $L=25, 50 \sigma$, average width $h_0=10,12\sigma$ and different values of the modulation of the channel; here $\sigma$ is the monomer size, taken as the unit length.
The corrugated channel is characterized by three characteristic length scales: the minimum and maximum widths $h_{min}$ and $h_{max}$ and the corrugation length $L$ along the channel axis (see Fig.~\ref{fig:model}). One usually recasts $h_{min}$ and $h_{max}$ into the entropic barrier $\Delta S = 2 \ln(h_{max}/h_{min})$ and the average width $h_0 = (h_{min}+h_{max})/2$.\\  
Remarkably, Fig.\ref{fig:free_energy_comparison}b) shows that for tailored channel geometries, {upon growing the polymer length $N$ the diffusion coefficient firstly decreases with $N$, it attains a minimum and then it grows again. This behavior,} differs from the expected $1/N$ Rouse behavior (i.e., upon neglecting hydrodynamic interactions). 

In order to clarify the physical origin of such a phenomenon, we derive an analytical model that maps the $3D$ dynamics of the confined polymers into the dynamics of a point particles moving in an effective $1D$ potential. In order to do so, we follow the Fick-Jacobs approximation~\cite{Zwanzig,Reguera2006,Dagdug2012_2,Dagdug2015,Malgaretti2013} that has been validated and used to study the dynamics of diverse confined systems (see Ref.\cite{Malgaretti2023} and references therein for a brief Review). 
A similar model has been derived~\cite{bianco2016} (see Suppl. Mat. 1A for a brief summary) under the condition that the 
polymer gyration radius is much smaller than the channel period, $L$. If so, naming $x$ the position of the centre of mass of the polymer along the channel axis, it is possible to model the effect of the confinement as a local contribution to polymer's free energy~\cite{bianco2016}. 
However, upon increasing the degree of polymerization $N$, such assumption is bound to fail. The polymer is, at some point, large enough to experience multiple channel periods at any location of the center of mass $x$, as sketched in Fig.~\ref{fig:model}. 
Hereby, we propose  an approach to construct an effective free energy for arbitrarily long polymers. 
The free energy should account for the diverse confining scenarios experienced along the chain. Accordingly, we integrate the local free energy from Ref.~\cite{bianco2016} over an interval  equal to the average magnitude of the end-to-end vector ($R_e$) and centered at the location of the center of mass
\begin{equation}
    \beta {F}(x) = \frac{1}{R_e} \int\limits_{x-R_e/2}^{x+R_e/2}{\beta {F}_0(x')dx'} 
    \label{eq:newf}
\end{equation}
where $\beta {F}_0(x)$ is the polymer free energy from Ref.\cite{bianco2016} and $R_{e}$ depends on $N$ and on the confinement (see Suppl. Mat. 1B). Finally, using Eq.~\eqref{eq:newf}, we compute the long time diffusion coefficient via the Fick-Jacobs formula~\cite{Zwanzig,Reguera2006,Dagdug2012_2,Dagdug2015,Malgaretti2023} 
\begin{equation}
    \frac{D}{D_N}=\frac{1}{(1+\Gamma) \langle e^{-\beta {F}(x)} \rangle_x \langle e^{\beta {F}(x) } \rangle_x} 
    \label{eq:D_fj}
\end{equation}
where $\Gamma = 2 (h_{max}-h_{min})^2/L^2$ is the so-called Zwanzig coefficient~\cite{Zwanzig}; for most cases considered in this paper $\Gamma$ is small ($\Gamma \approx 10^{-2}$) and can be rather safely ignored.

First, we compare theoretical and numerical free energies in Fig.~\ref{fig:free_energy_comparison}a as a function of the coordinate $x$, that again marks the position of the centre of mass of the polymer along the longitudinal axis of the channel, for $L= 50\sigma$, $\Delta S=$0.81 and $h_0=12\sigma$. As shown, 
the comparison is very favourable up to polymers with $N \approx$ 125. Afterwards, it becomes less quantitative, even though the free energy difference $\beta \Delta F$ is quite well captured by the model, even at $N=200$.\\
In Fig.~\ref{fig:free_energy_comparison}b we compare theoretical and numerical data for the long-time diffusion coefficient $D/D_0$, where $D_0$ is the diffusion coefficient of a single monomer, as a function of $N$ for different values of $L$, $h_0$ and $\Delta S$. Theoretical data are computed using Eq.~(\ref{eq:newf}) and (\ref{eq:D_fj}); numerical data are reported only if the vast majority ($>90\%$) of the trajectories diffused at least up to  one channel corrugation from their initial point $x_0$ (i.e. up to $x_0 \pm L$). (see Suppl. Mat. 2D).\\
We find a good agreement between  theoretical and numerical data, both showing a clear non-monotonic behaviour. 
Remarkably, the position of the diffusion minimum $N_{min}$ is always rather well captured by the theory. 
The non-monotonic nature of the diffusion coefficient can be emphasized looking at $D/D_N$ (see Suppl. Mat. 2B), $D_N$ being the bulk diffusion coefficient of a linear polymer of $N$ monomers. Such a quantity highlights the effect of the confinement, removing the contribution due to the increase in size. However, we highlight in Fig.~\ref{fig:free_energy_comparison}b that the diffusion coefficient may vary by one or even two orders of magnitude, upon changing the polymer size roughly by a factor of two. Thus, the difference in the diffusion time scales for polymers of slightly different size can be substantial: such a difference can easily be measured in experiments or exploited for material design.

In many of biological scenarios, a relevant quantity is the time at which the polymer crosses a bottleneck for the first time i.e. its first passage time. Typically, the mean of the first passage time distribution is take as representative of the "typical time" taken by the polymer to cross a barrier. However, in several systems~\cite{d,e,g,Malgaretti2019} the distribution of the first passage time is quite broad and skewed, thus its mean may not be so significant. Accordingly, we employ Eq.~(\ref{eq:newf}) to compute the Mean First Passage Time (MFPT) $T_1$, its variance $\sigma_{T_1}$ and, the so-called coefficient of variation $\gamma= \sigma_{T_1}/T_1$~\cite{d,e,g,Malgaretti2019} that quantify the statistical likelihood of a departure from the mean.  
Following\cite{Malgaretti2019}, we define the Mean First Passage Time as the time required to the centre of mass of the polymer to displace a distance $L$ in either directions along the channel axis, from its initial point. From Eq.~\eqref{eq:newf}, the MFPT can be computed as
\begin{equation}
    T_1(x_0) = \frac{1}{D_N} \int \limits_{x_0}^{x_0+L} {dx' e^{\beta {F}(x')}} \int \limits_{x_0}^{x'}{dx'' e^{-\beta {F}(x'')} }
\end{equation}
where $x_0$ is the initial point of the trajectory (see Suppl. Mat. 1E, 2D).

\begin{figure}[t]
\centering
 \includegraphics[width=0.40\textwidth]{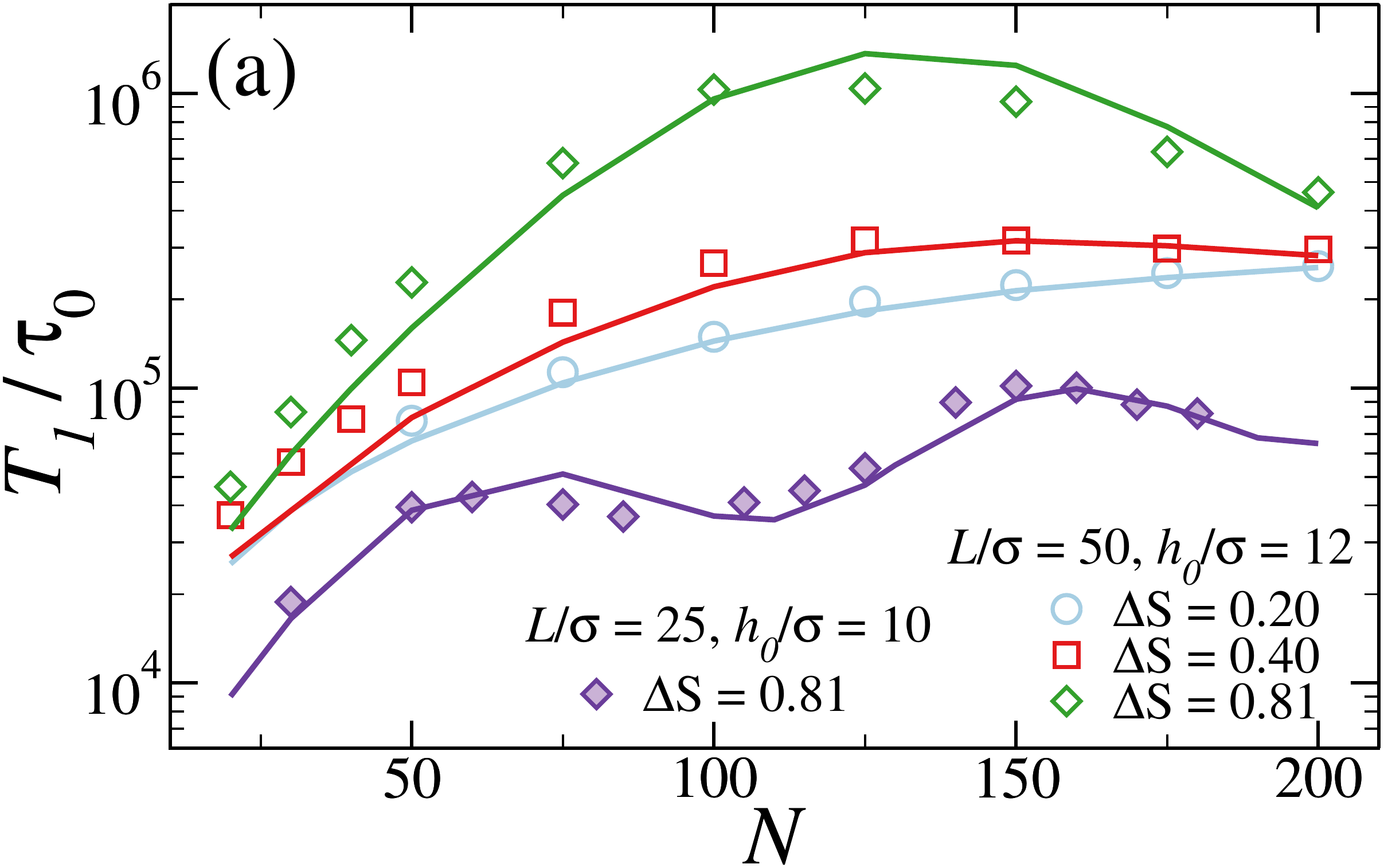}
 \includegraphics[width=0.40\textwidth]{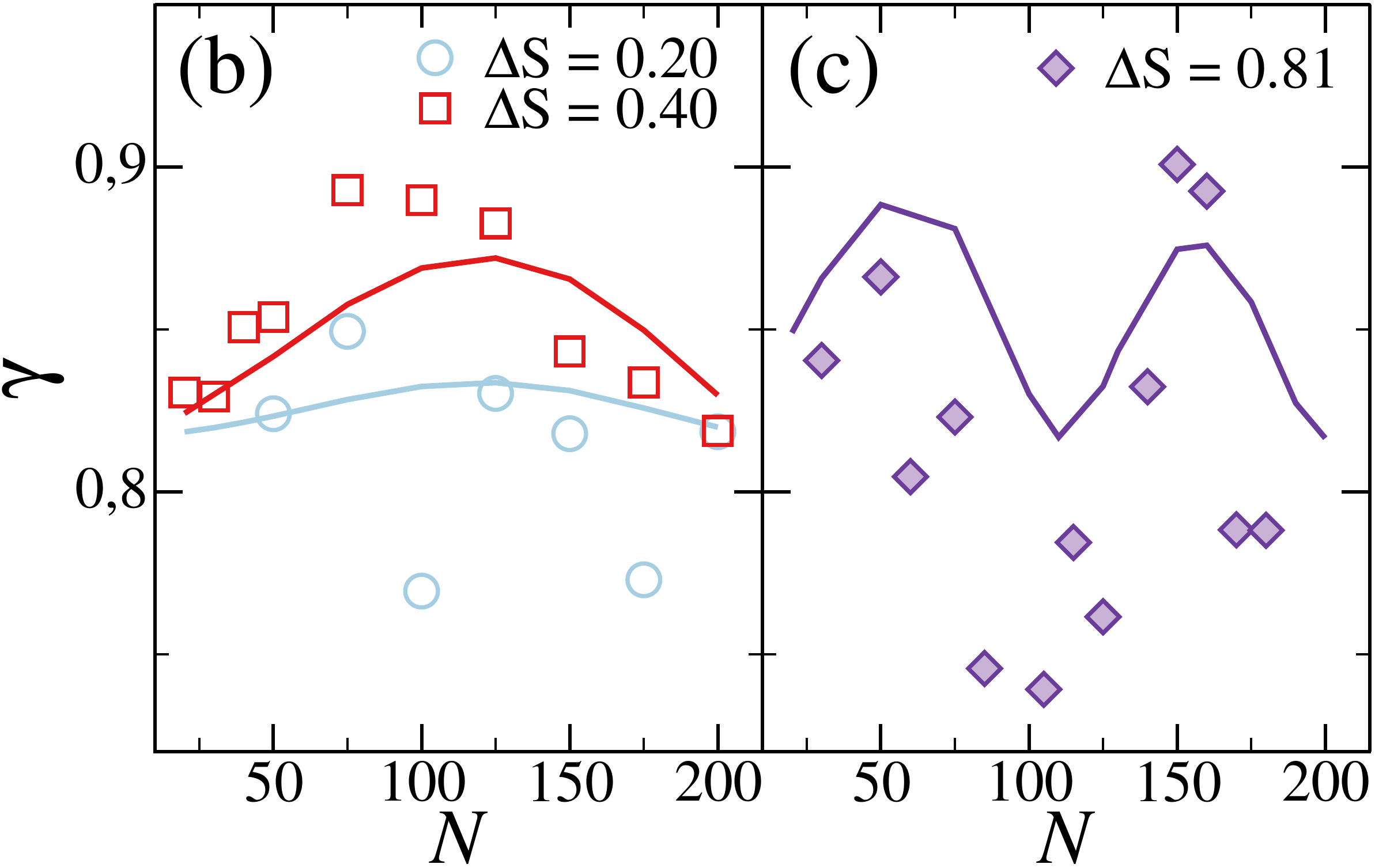}
\caption{(a) Mean Passage Time $T_1\tau_0$, normalized over time unit $\tau_0$, and (b),(c) coefficient of variation $\gamma$ as a function of $N$ for different values of $\Delta S$, $h_0$ and $L$. Symbols refer to results from simulations, lines to theoretical calculations.}
 \label{fig:T1_comparison}
\end{figure}

We report, in Fig.~\ref{fig:T1_comparison}, the comparison between numerical and analytical results for both $T_1$ and $\gamma$. 
In particular, we observe in Fig.~\ref{fig:T1_comparison}(a) that the comparison for the MFPT is again quantitative; the comparison for $\gamma$  (Fig.~\ref{fig:T1_comparison} b and c) is instead mostly qualitative. This has to be expected: for $\gamma \approx 1$ the standard deviation is comparable to the mean and a considerable amount of data is required for a precise estimation of $\sigma_{T_1}$. Nevertheless, theory and simulations agree on the range of values of $\gamma$ as well as on the qualitative trends. 

Such a remarkable agreement shows that the simple model in Eq.~(\ref{eq:newf}) not only provides quantitatively reliable predictions for average quantities such as the diffusion coefficient and the MFPT, but it is also reliable for what concerns higher moments of the first passage time distribution.

\begin{figure}[t]
\centering
 \includegraphics[width=0.4\textwidth]{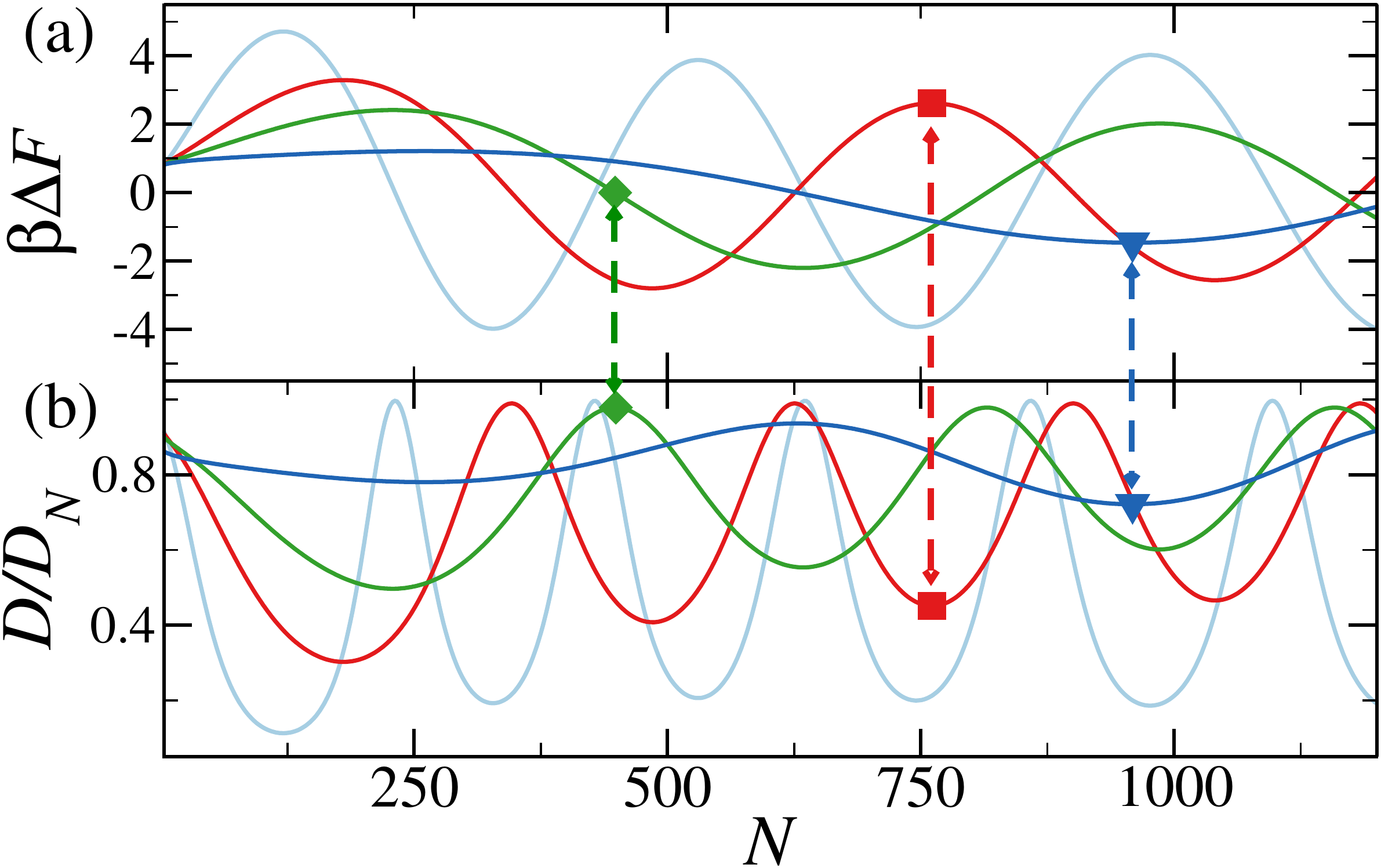}
\caption{(a) Theoretical free energy difference $\beta \Delta F$ and (b) theoretical diffusion coefficient $D/D_N$ as a function of the degree of polymerization $N$ for a channel of length $L = 50\sigma$, $\Delta S=$0.81 and $h_0 =12\sigma$ (light blue), $h_0 =20\sigma$ (red), $h_0 =28\sigma$ (green), $h_0 =48\sigma$ (blue). Symbols refer to the position of (a) the extrema and the zeros of $\Delta F$ (b) maxima and minima of $D$.}
 \label{fig:DF_D_theo}
\end{figure}

This motivated us to examine the predictions of the theory for a broad range of  $L$, $h_0$ and $\Delta S$ values, exploiting the predictive power of Eq.~(\ref{eq:newf}). It is instructive to look at the theoretical results as a function of $N$. In Fig.~\ref{fig:DF_D_theo} we plot the theoretical estimates for $\beta \Delta F$ (Fig.~\ref{fig:DF_D_theo}a) and for $D/D_N$ (Fig.~\ref{fig:DF_D_theo}b). The theory predicts a periodic dependence for both quantities on the polymer size $N$, as also observed in our simulation data and in the literature\cite{marenda2017sorting}. The period of the oscillations as well as that the ratio between the diffusion coefficient at the extrema, $D_{max}/D_{min}$, varies with the channel average section $h_0$. 
As expected, the theoretical approach identifies $\beta \Delta F$ as the ``driving force'' for the non-monotonic diffusion: indeed, the frequency of the oscillation of $\beta \Delta F$ is always half the frequency of the oscillation of $D/D_{N}$. As shown in Fig.~\ref{fig:DF_D_theo}, the extrema of $\beta \Delta F$ correspond to the minima of $D/D_{N}$, while the zeros of $\beta \Delta F$, indicating a flat free energy landscape, correspond to the maxima of the diffusion. 

\begin{figure}[t]
  \includegraphics[width=0.45\textwidth]{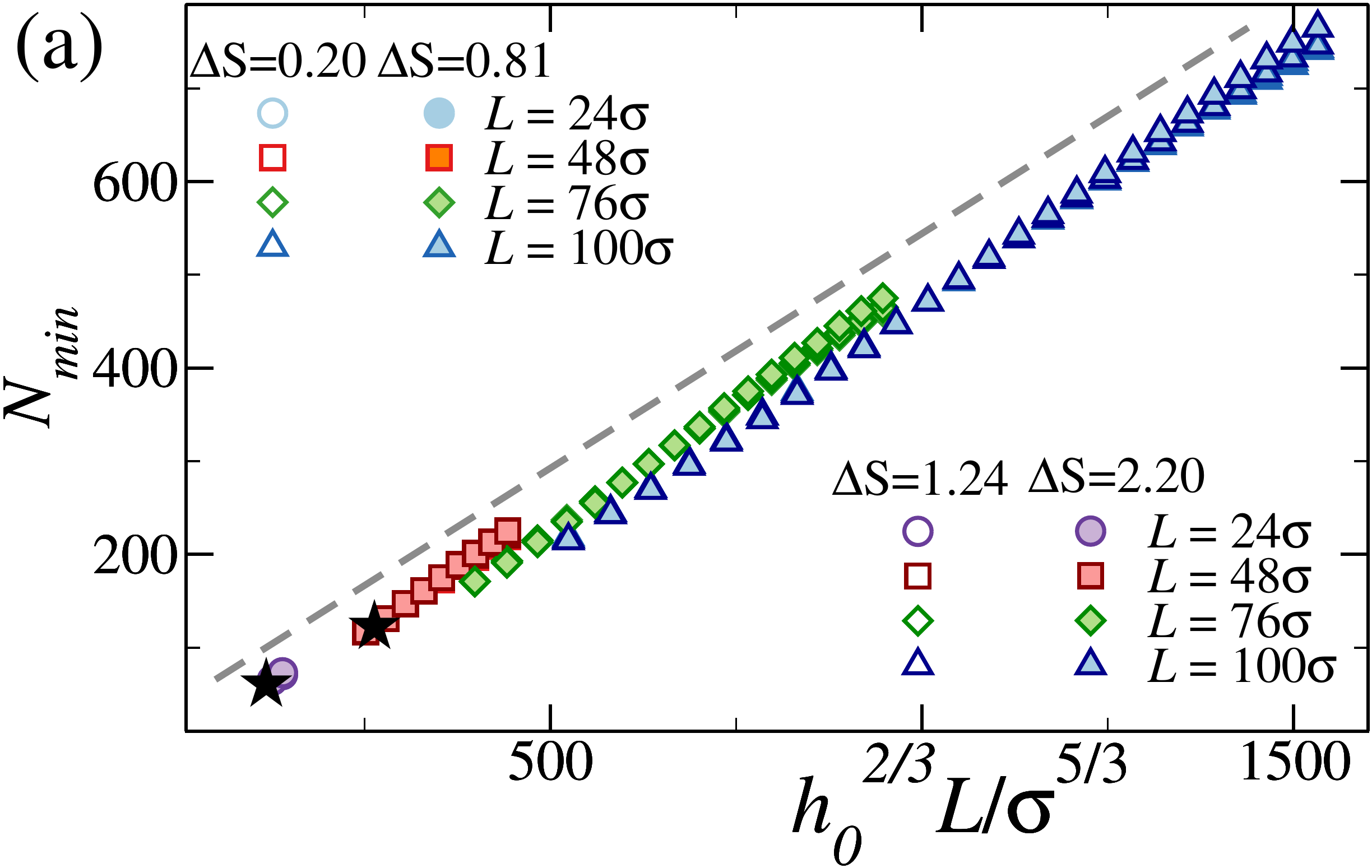}
  \includegraphics[width=0.45\textwidth]{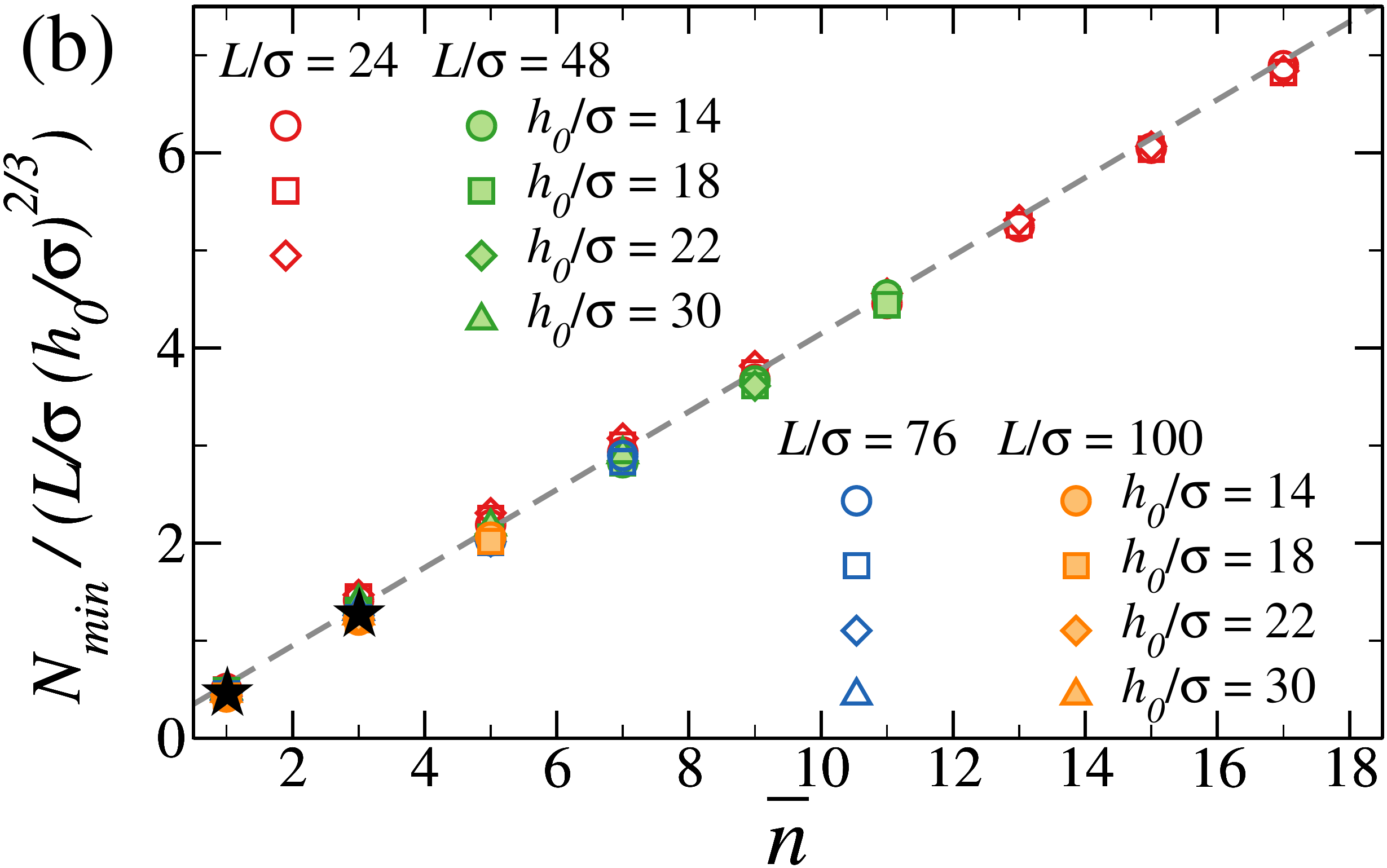}\\
   \caption{(a) Scaling of the position of the first minimum of the diffusion coefficient as a function $h_0^{2/3} L/\sigma^{5/3}$ for different values of $L$ and $\Delta S$.
    (b) Scaling of the positions of the diffusion minima $N_{min}$ as a function of their appearance index for different values of $L$ and $h_0$ at $\Delta S =$ 0.81. The black star symbols refer to the numerical data, extracted from Fig.~\ref{fig:free_energy_comparison}b.
    }
  \label{fig:scalings}
\end{figure}

Finally, we focus on the theoretical prediction of the position of the diffusion minima. Fig.~\ref{fig:scalings}a shows the scaling properties of the first minimum of the diffusion coefficient as function of the channel width. Remarkably, the data obtained at different values of $\Delta S$ and $L$ collapse onto a master curve, when reporting $N_{min}$ as a function of $h_0^{2/3}L/\sigma^{5/3}$. Considering all the subsequent minima $N_{min}(\bar{n})$ in Fig.~\ref{fig:scalings}b, ranked by their appearance index, for $\Delta S=0.81$ and different values of $L$ and $h_0$, we can clearly observe a linear behaviour. Both results can be understood using a blob model~\cite{DeGennes_book} (see Appendix A), that predicts { the scaling reported above; indeed, }the blob model properly captures not only the position of the first minimum but also that of all the other minima ( Fig.~\ref{fig:scalings}b). 
It also suggests that the position of the minimum (and, actually, of all minima) is  independent on $\Delta S$. Such a finding is in agreement with the free energy model (see Appendix B). 

The blob model can inspire  alternative approaches to an  effective free energy. If $h_0$ is the blob length scale, then one may introduce non-local contributions to the free energy via an effective channel width, averaging the channel profile over the length scale $R_e$. Albeit less precise, this alternative approach still maintains a very good agreement with numerical data (see Appendix B). This shows that the idea of incorporating contributions by a suitable averaging is robust.

In summary, {we have shown that the diffusion coefficient of linear polymers across varying-section channels has a non-monotonous dependence on the polymer length $N$. In particular, 
we observe that the deviation from the Rouse behaviour can be significant ($10-100-$fold) and can be exploited to design devices aiming at polymer sorting.  In order to understand such a counter-intuitive behavior we have derived (and validated against numerical data) an effective free energy approach which captures the non-monotonic behavior quantitatively. 
The approach, based on the Fick-Jacobs theory, incorporates non-local effects of the confining channel through the average of the local free energy. Our model shows that the non-monotonous (and oscillating) behavior is due to the (periodic) smearing out of the effective free energy barrier for polymers that occupy an integer number of channels (see Fig.~\ref{fig:DF_D_theo}). Remarkably, a similar behavior has been observed experimentally, for example in the case of polymer transported across an array of entropic traps~\cite{Kim2017} as well as for pluronic gels~\cite{You2017}.} 
Moreover,  
in many circumstances, the dynamics is controlled by the first passage of polymers across a pore. 
We have used our model to predict both the mean and the variance of the first passage time distribution. As shown in Fig.~\ref{fig:T1_comparison} our model precisely predicts the mean and qualitative captures the dependence of the variance of the first passage distribution on $N$.

Finally, we exploited the remarkable agreement between theoretical and numerical data to investigate the scaling of the minima of the diffusion coefficient as a function of the channel geometry and we have captured such scaling via blob theory.    
{Even though our model and numerical results do not account for finite bending rigidity, this can be effectively accounted for by renormalizing the monomer size in the model to the persistence length and the number of monomers to the number of Kuhn segments~\cite{doi_edwards}.
Accordingly, we expect our results to hold in presence of a moderate bending rigidity and deviation can be expected when the persistence length becomes comparable with the full contour length of the polymer as well as for rigid rods~\cite{Malgaretti2021}.}
Our findings present a significant theoretical improvement in the understanding of the diffusion of polymers in complex landscapes and, further, provide very useful tools for characterizing the diffusion properties in complex porous materials. One could, for example, envisage a "sponge", designed to have pores of different roughness that can effectively sort different  polymers by size in a passive way. Such a device could be useful in situation where it is hard to produce a flow or where the polymers may be degraded by the action of an external force. Finally, from a theoretical perspective, it would be interesting to check if this effective approach works far from equilibrium, for example with active polymers. Work is in progress in that respect, with the aim of unravelling the effect of activity on the polymer dynamics in corrugated channels.

\begin{acknowledgments}
The authors thank E. Carlon for helpful discussions. E.L. acknowledge the contribution of the COST Action CA17139. V. B. acknowledges the support the European Commission through the Marie Skłodowska–Curie Fellowship No. 748170 ProFrost. E.L. acknowledges support from the MIUR grant Rita Levi Montalcini and from the HPC-Europa3 program. C.V acknowledges funding from MINECO grants EUR2021-122001, PID2019-105343GB-I00, IHRC22/00002. The computational results presented have been achieved using the Vienna Scientific Cluster (VSC) and the Barcelona Supercomputing Center (BSC-Marenostrum).
\end{acknowledgments}

\section{Appendix A: Blob model}
\label{sec:blob}

We describe here the blob model, employed to rationalize the scaling of the diffusion minima reported in the main text. We assume that the polymer organizes into blobs of radius $\sim h_{0}$/2. A polymer made of $N$ monomers has $n_b = N/M$ blobs, where $M$ is the  number of monomers in a blob, given by $\alpha M^{\frac{3}{5}}= h_{0}/(2\sigma$), $\alpha$ being a dimensionless constant{, and $\sigma$ the monomer size}. It follows that 
\begin{equation}
M=\left(\frac{h_{0}}{2\alpha\sigma}\right)^{\frac{5}{3}}
\end{equation}
In blob theory, the length of a polymer $\ell$ scales linearly with the number of blobs; in our case $\ell = n_b h_0$. 
We argue that, when $\ell$ exactly matches an integer multiple of $L$ $\ell = \bar{n}L$ ($\bar{n}=1,2,3,...$) we expect $\beta \Delta F$ to be null, i.e. $\beta F(x)$ to be constant.
In fact, for such a case, the polymer experiences the "same confinement" (in the sense of the effective free energy) irrespective of the position of the center of mass, as a consequence of the 
{periodicity} of the corrugation profile. Since this is true only for the set of values $\ell = \bar{n}L$ ($\bar{n}=1,2,3,...$), 
this also entails that $\beta \Delta F$ as a function of $N$ is oscillating; the extremal point will be then placed at  $\ell = \bar{n} L/2$. 
As seen in the main text, the extrema in the free energy difference correspond to minima in the diffusion: thus 
\begin{equation}
\frac{\bar{n}}{2}L = n_b^{min} h_0 = \frac{N_{min}}{M}h_{0}
\end{equation}
sets the condition for $N_{min}$. Accordingly, we get
\begin{equation}
N_{min} = \frac{\bar{n}}{2}\frac{L}{h_{0}}\left(\frac{h_{0}}{2\alpha\sigma}\right)^{\frac{5}{3}}=\frac{\bar{n}}{2}\frac{1}{\alpha^{\frac{3}{5}}}\frac{L}{\sigma}\left(\frac{h_{0}}{2\sigma}\right)^{\frac{2}{3}}
\end{equation}
As noticed in the main text, this relation can be recast as
\begin{equation}
    \frac{N_{min}^{3/2}}{(L/\sigma)^{5/2}} \propto h_0/L
\end{equation}
which highlights the two length scales of the system $h_0$ and $L$. As noticed in the main text, this relation is unaware of $\Delta S$, which is also reflected by the numerical data (see Suppl. Mat. Section 1.E). Finally, it is interesting to notice that, {via our model,} 
a similar scaling is {expected also for the maxima of the diffusion coefficient.} 

\section{Appendix B: Effective free energy approaches}
\label{sec:eff_approaches}
We consider three different approaches to construct an effective free energy; each one is characterized by a different approximation. While all of them qualitatively explain the non-monotonicity of the diffusion coefficient as a function of $N$, the agreement with numerical data varies quantitatively.\\
We start recalling the definition of the free enrgy in the case of a polymer whose gyration radius is much shorter then the channel length 
\begin{align}
    \beta {F}_0(x) =& -(d-1) \Bigg\{ \ln \left[ \frac{16 h(x)}{h_0 \pi^2}\right]\nonumber \\
    &  + \ln \left[ \sum_{p=1,3,..}^{\infty}\frac{1}{p^2} \exp \left( - \pi^2 p^2 \left( \frac{R_g}{2h(x)} \right)^{1/\nu} \right) \right] \Bigg\}
    \label{eq:jcp_result}
\end{align}
where the values of $\nu$ and $R_g$ depend on the polymer chain considered (Gaussian or self-avoiding). For our purposes, we used the following functional form for the bulk gyration radius $R_g^{b} = 0.58735 \cdot N^{0.588} \cdot (1-0.435588 \cdot N^{-0.2228})$, with thus $\nu=0.588$.\\
The \textit{first approach} is the one reported in the main text: we average the free energy
Eq.(\ref{eq:jcp_result}) over the characteristic length scale given by the magnitude of the (average) end-to-end vector. This approach provides the best comparison with the numerical data. The main idea is to include non-local contributions to the free energy. These contributions come from the fact that a sufficiently large polymer is, at any given time, extended along the channel and experiences different degrees of confinement. The resulting free energy reads
\begin{equation}
    \beta {F}(x) = \frac{1}{R_e} \int_{x-R_e/2}^{x+R_e/2}{\beta {F}_0(x')dx'} 
    \label{eq:newf_suppl}
\end{equation}
where $\beta {F}_0(x')$ is the polymer free energy from Ref.\cite{bianco2016} and $R_{e}$ depends on $N$ and on the confinement.
\\The \textit{second approach} aims at incorporating the hypothesis of the blob model (see Appendix A) into the calculation of the free energy. The blob model assumes that the polymer organizes into section{s} with correlation length (blob radius) $h_0$; thus $h_0$ is the relevant length scale of the confinement. Yet, for a single blob picture, such as the one proposed here, an effective confinement length $h_0^*$ is necessary; it has to include the contributions of non-locality and of the channel geometry. We thus define $h_0^*(x)$ as
\begin{equation}
    h_0^*(x) = {\frac{1}{R_e}}\int_{x-R_e/2}^{x+R_e/2}{h(x') dx'}
    \label{eq:h0eff}
\end{equation}
The free energy is given by Eq.~(\ref{eq:jcp_result}) where $h(x)$ is replaced with $h_0^*(x)$.\\
Finally, the \textit{third approach} aims at introducing a minimal perspective. Indeed, we compute the free energy difference using the first approach but we drastically simplify the functional form by considering a piece-wise linear function 
\begin{equation}
    \beta {F}(x) = 
    \begin{cases}
    -\frac{2 |\beta \Delta {F} | x}{L} &\qquad x < L/2 \\
    -2 |\beta \Delta {F} | + \frac{2 |\beta \Delta {F} | x}{L} &\qquad x \ge L/2
    \end{cases}
\end{equation}


\begin{figure}[h!]
\centering
 \includegraphics[width=0.35\textwidth]{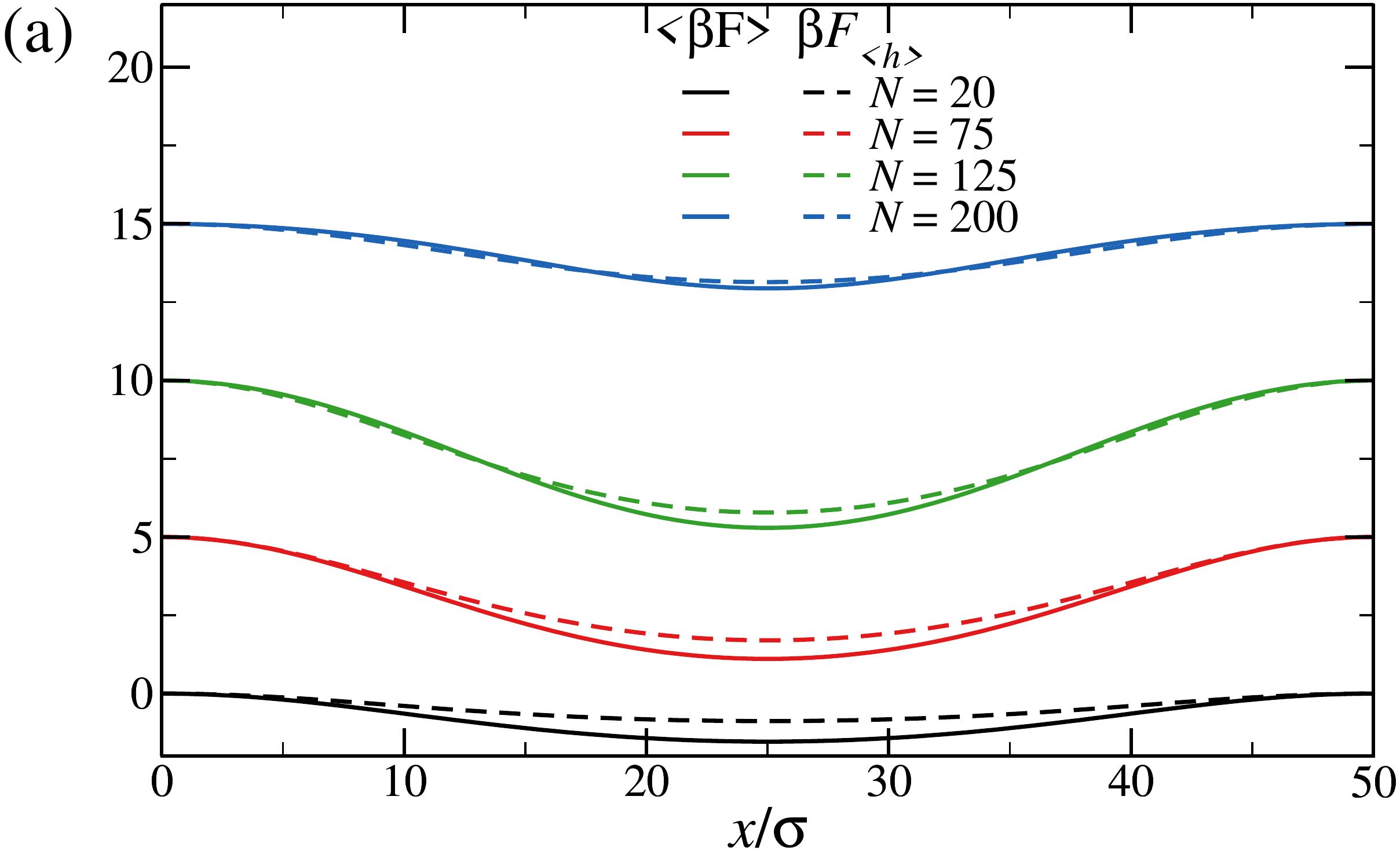}
 \includegraphics[width=0.35\textwidth]{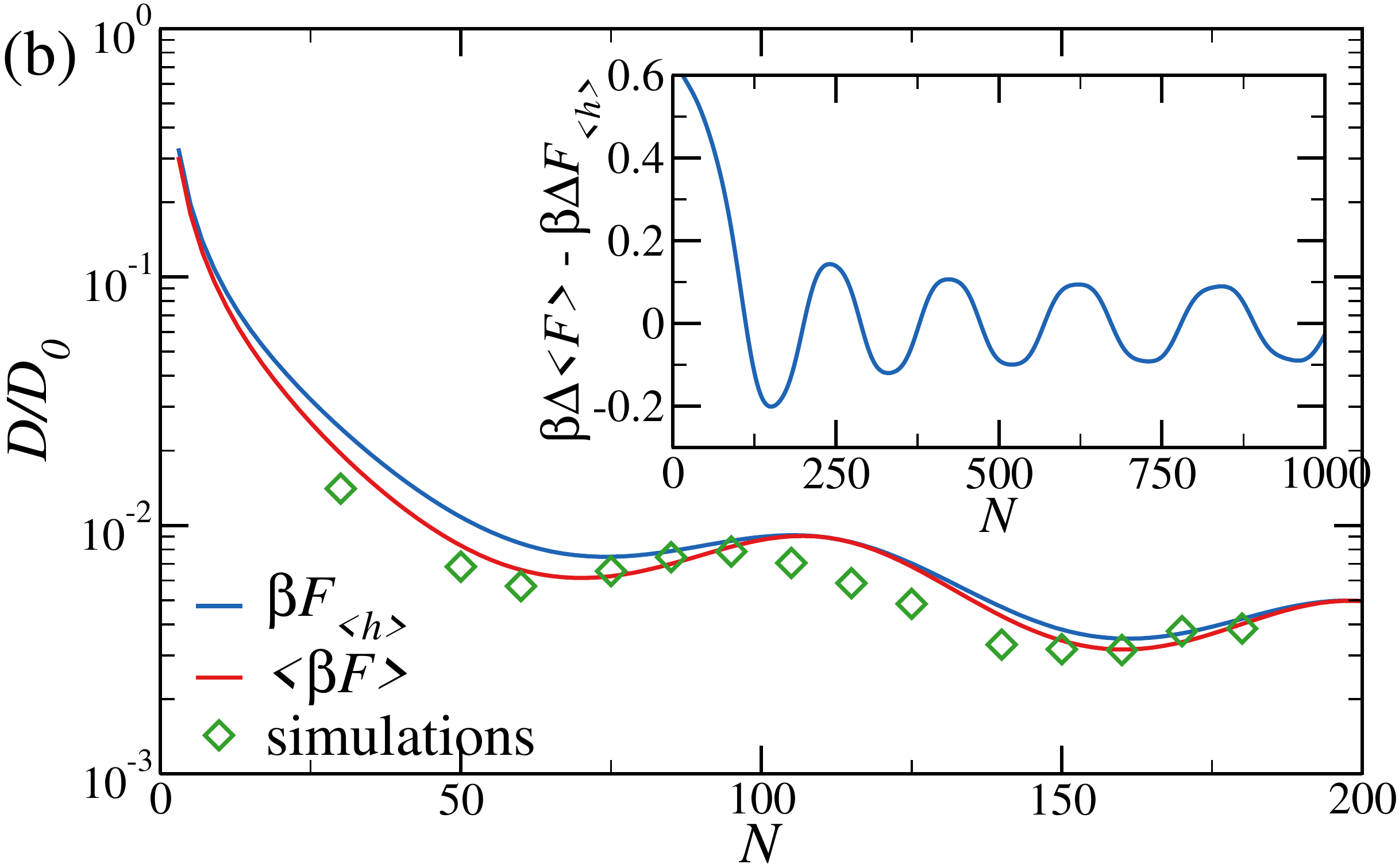}
\caption{(a) Comparison between the free energy from the first approach  Eq.~(\ref{eq:newf_suppl}) (solid lines) and the second approach Eqs.~(\ref{eq:jcp_result}) and (\ref{eq:h0eff}) (dashed lines), as a function of the channel coordinate. Curves have been shifted arbitrarily. (b) Comparison between the prediction of the first two theoretical approaches and numerical data as a function of $N$. Inset: Difference between the free energy differences obtained through the first and second theoretical approaches as a function of $N$. In all panels $\Delta S=0.81$ while panel (a): $L=50\sigma$, $h_0=12\sigma$; panel (b) $L=25\sigma$, $h_0=10\sigma$ }
 \label{fig:compare_theory}
\end{figure}
In Fig.~\ref{fig:compare_theory}, we report the comparison between the first and second approach. The first approach incorporates non-local effect averaging the free energy, the second approach averaging of the corrugation profile. In Fig.~\ref{fig:compare_theory}(a) we compare the theoretical free energies for different values of $N$ for the same systems considered in the main text, i.e. for $\Delta S=0.81$, $L=50\sigma$, $h_0=12\sigma$. The comparison is favourable and, upon increasing $N$ the two approaches produce the same free energy difference. Indeed, as reported in Fig.~\ref{fig:compare_theory}(b), the diffusion coefficient predicted by the two approaches are very similar (here for $\Delta S=0.81$, $L=25\sigma$, $h_0=10\sigma$). Albeit the second approach is slightly less precise, it is able to reproduce very well the important features of the numerical data, such as the position of the two diffusion minima. It also remains predictive even when the polymer size is larger then the corrugation length (see Suppl. Mat. 2.C). Further, in the inset of Fig.~\ref{fig:compare_theory}(b) we plot the difference between the free energy difference for the two approaces in a large span of values of $N$. Notably, the absolute difference never exceeds 0.6 $k_B T$ (0.1 $k_B T$ if $N>200$).  

\begin{figure}[h!]
\centering
 \includegraphics[width=0.35\textwidth]{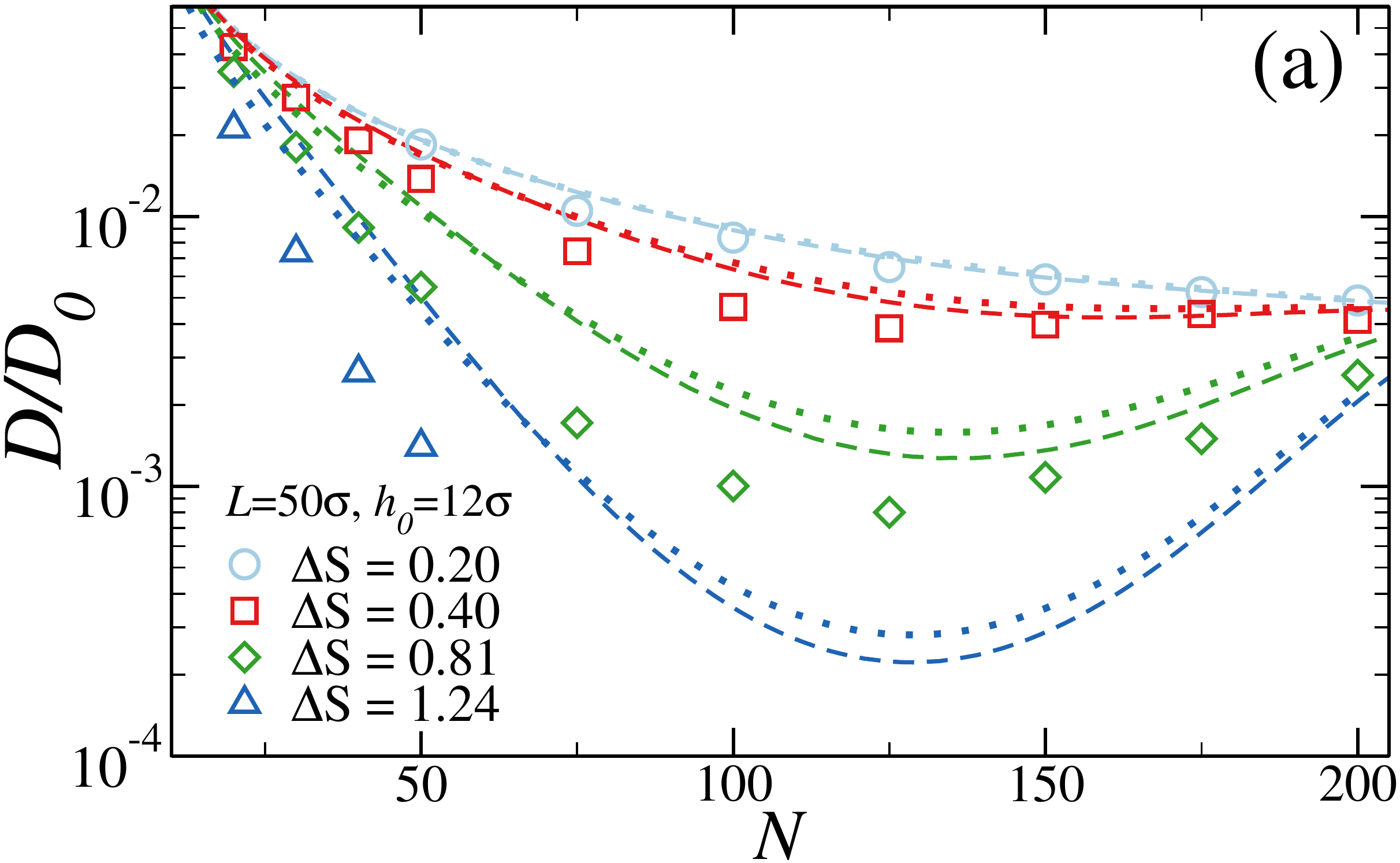}
 \includegraphics[width=0.35\textwidth]{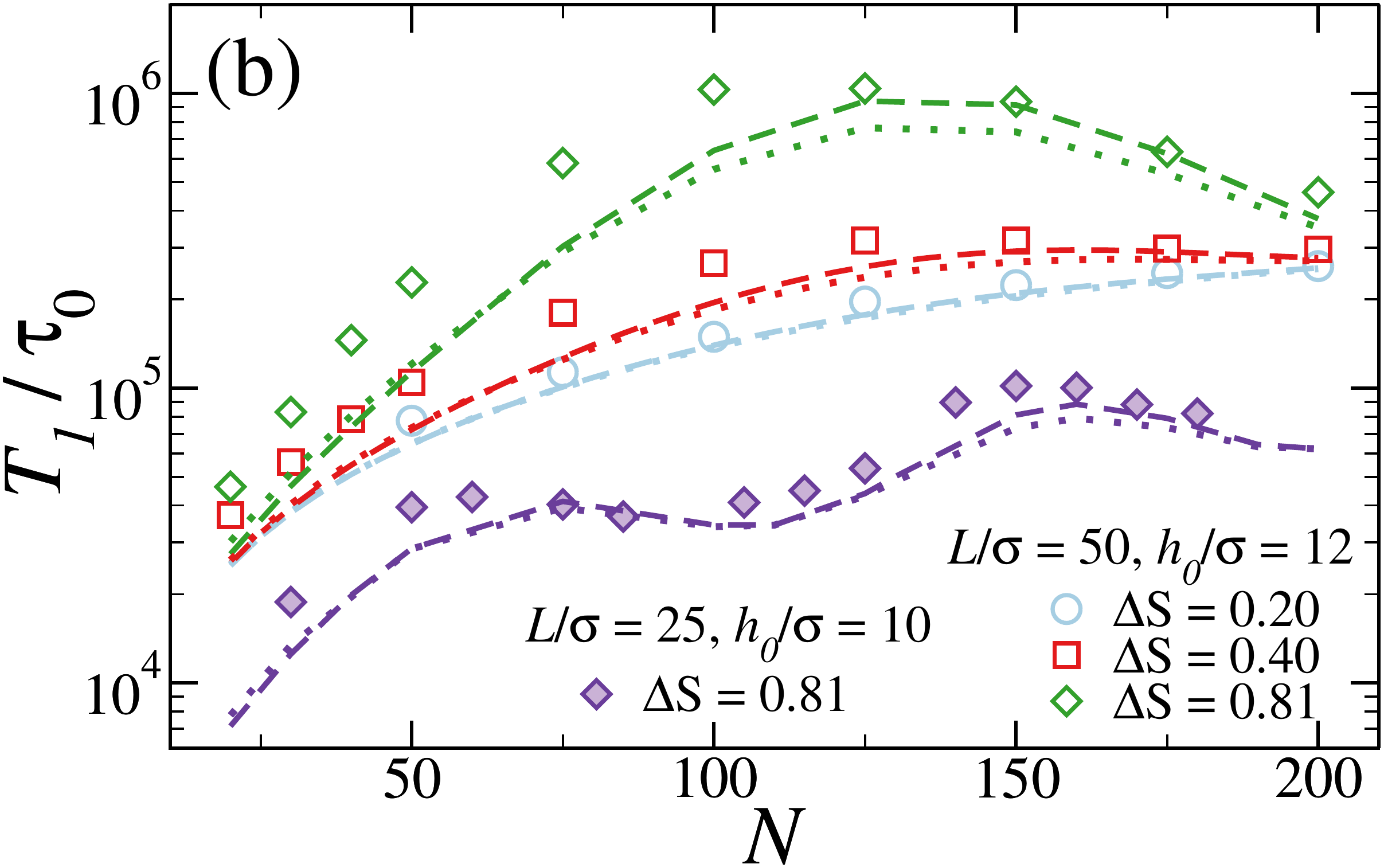}
\caption{Comparison between the alternative theoretical approaches and the numerical data for (a) the diffusion coefficient and (b) the Mean Passage Time, for different values of $\Delta S$, $L$ and $h_0$.}
 \label{fig:compare_theory2}
\end{figure}
Finally, we compare the second and third approaches with numerical data. In Fig.~\ref{fig:compare_theory2}(a) we compare the diffusion coefficient while in Fig.~\ref{fig:compare_theory2}(b) we compare the Mean Passage Times. In both cases we observe that the alternative approaches are less precise then the free energy average but retain the salient features, such as the position of the minimum, and provide a decent comparison with numerical data, considering there are no free parameters and no input from the simulations.    

\providecommand{\noopsort}[1]{}\providecommand{\singleletter}[1]{#1}%
%


\pagebreak
\onecolumngrid

\appendix

\title{The bigger the faster: non-monotonous translocation time of polymers across pores - Supplemental Material}

\section{Theoretical approaches}

\subsection{Free energy of a linear chain in a corrugated channel}

We consider a linear polymer in equilibrium, confined in a channel of characteristic length $L$, corrugated with a  profile
\begin{equation}
    h(x) = \frac{h_0}{2} + \frac{h_{min}-h_{max}}{2} \cos{\left( \frac{2 \pi x}{L}\right)}
\label{eq:channel}
\end{equation}
with $h_{max}$ and $h_{min}$ being the width of the channel at the widest and thinnest point, respectively, and $h_0 = (h_{min}+h_{max})/2$. The free energy of such polymer is reported in Ref.\cite{bianco2016}. The result was developed within the Fick-Jacobs approximation and it is based on two main assumptions:
\begin{enumerate}
    \item the dynamics of the polymer is fast enough that the chain is able to relax completely in between successive translocations, i.e. moving through a bottleneck to a neighbour corrugation. This allows for uncorrelated translocations. We impose that the polymer diffusion time should be larger than the slowest relaxation time of the chain; this, in terms of the gyration radius of the chain $R_g$, reads $R_g \ll \pi L/\sqrt{2}$
    \item the amplitude of the channel is varying slowly $\frac{dh(x)}{dx} << 1$; at fixed entropic barrier, one can equivalently impose that the average width of the channel should be smaller than its length $h_0 < L$
\end{enumerate}
The free energy, in absence of an external force, reads
\begin{equation}
    \beta {F}(x) = -(d-1) \Bigg\{ \ln \left[ \frac{16 h(x)}{h_0 \pi^2}\right] + \ln \left[ \sum_{p=1,3,..}^{\infty}\frac{1}{p^2} \exp \left( - \pi^2 p^2 \left( \frac{R_g}{2h(x)} \right)^{1/\nu} \right) \right] \Bigg\}
    \label{eq:jcp_result_s}
\end{equation}
where the values of $\nu$ and $R_g$ depend on the polymer chain considered (Gaussian or self-avoiding). For our purposes, we used the following functional form for the bulk gyration radius $R_g^{b} = 0.58735 \cdot N^{0.588} \cdot (1-0.435588 \cdot N^{-0.2228})$, with  $\nu=0.588$.

\subsection{Reference values for $R_e$}
\label{sec:ref_ree}
The effective approach described in the main text is based on averaging Eq.~(\ref{eq:jcp_result_s}) over a certain length scale. Probably the most natural choice for such a length scale is the magnitude of the end-to-end vector $R_e$. Indeed, it is known that, upon increasing the degree of polymerization $N$, the gyration radius becomes a poor descriptor of the polymer size in a channel, as the chain becomes increasingly elongated. Further, we prefer $R_e$ to the so-called longitudinal extension of the chain $L_{\parallel}$, used in the description of semi-flexible chains under confinement, because i) $R_e$ is well defined in the bulk and ii) they are effectively very similar in magnitude, at least under sufficiently strong confinement. However, in a corrugated channel, the magnitude of the end-to-end vector is a function of the position of the centre of mass along the channel axis. Thus, for every position considered one should take the end-to-end vector at that position. Unfortunately, to the best of our knowledge, no expression for this quantity is available. As mentioned in the main text, a universal curve for the magnitude of the end-to-end vector of polymer chain confined in cylindrical channels with constant section was reported in the literature\cite{morrison2005shape}. We will thus approximate $R_e (x)$ in the corrugated channel with the (constant) $R_e$ computed in a channel of width $h_0$.

\begin{figure}[h!]
\centering
 \includegraphics[width=0.4\textwidth]{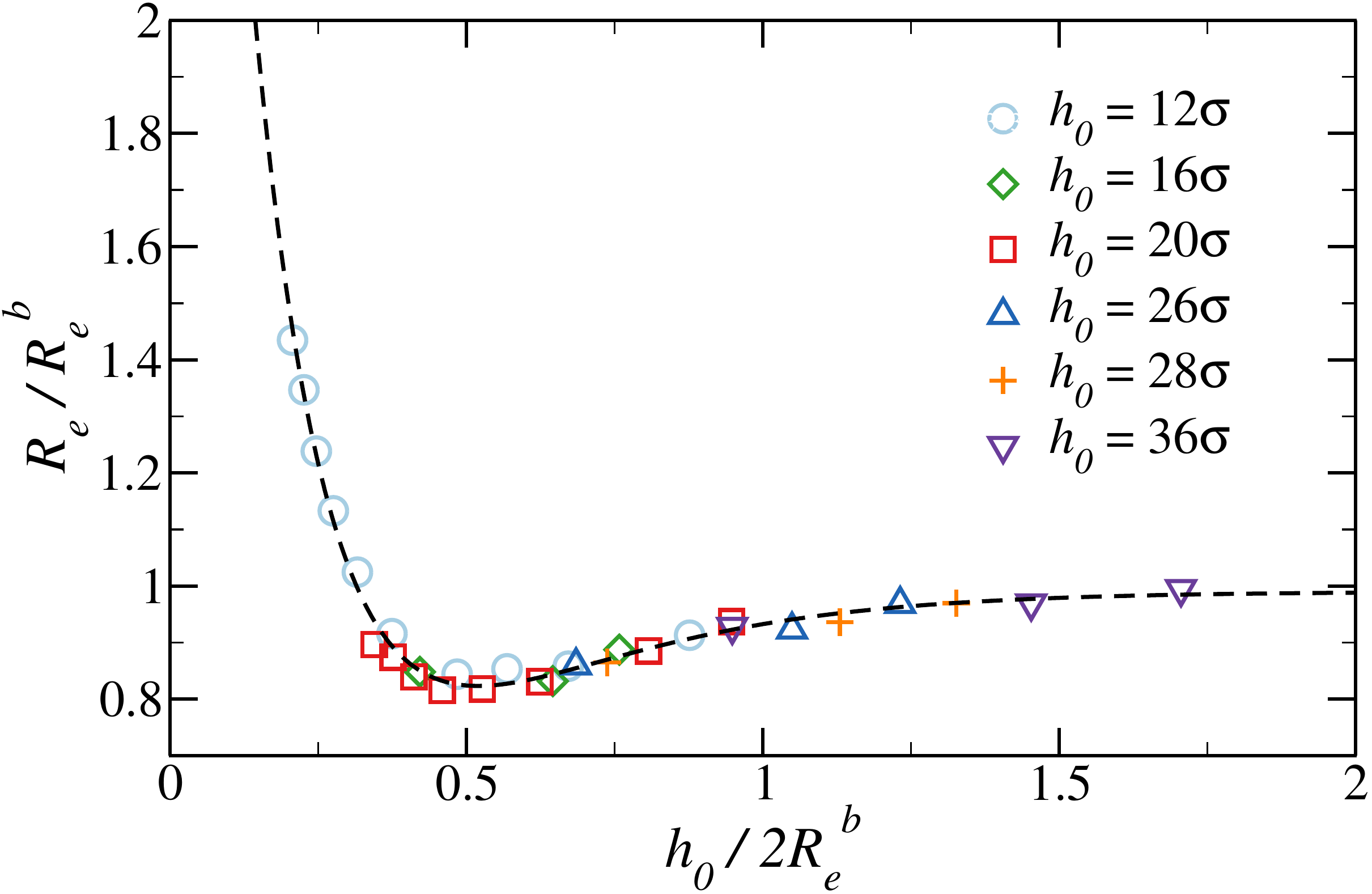}
 \includegraphics[width=0.4\textwidth]{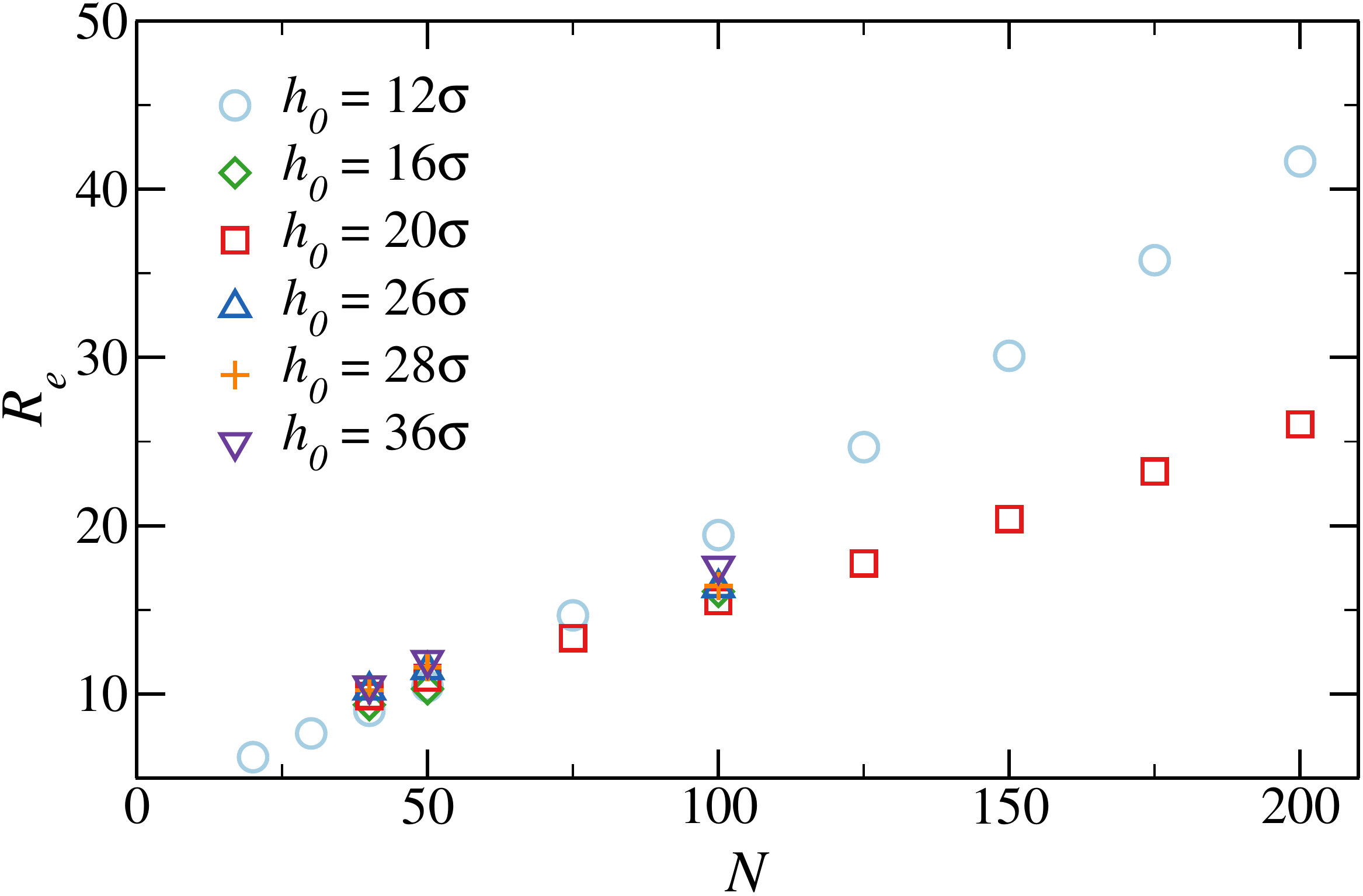}
\caption{(left) Master curve of $R_e/R_e^b$ as function of $h_0/(2 R_e^b)$. Symbols refer to results from numerical simulations, dashed line is an interpolation with a Morse-like function. (right) $R_e$ as a function of $N$ for different channel widths $h_0$.}
 \label{fig:Re_scaling}
\end{figure}

We performed simulations (see the section \ref{sec:sim_model} Simulation details) at different values of the channel section $h_0$ and  reproduced the master curve.
We then performed an interpolation to obtain the following functional form 
\begin{equation}
R_{e} = 0.166863 \cdot \left(1-\exp\left(-3.44827 \cdot \left(\frac{h_0}{2 R_{e}^b}-0.51978 \right) \right)\right)^2+0.823389    
\end{equation}
where $R_{e}^{b} = 1.43196 \cdot N^{0.588} \cdot (1-0.33039 \cdot N^{-0.2228})$ is the magnitude of the end-to-end vector in the bulk. 
Results of the simulations and of the interpolation are reported in Fig.~\ref{fig:Re_scaling}.

\subsection{Mean Passage Time}

We report here briefly how we compute the Mean Passage Time (MPT) $T_1$ and related quantities analytically. As explained in Ref.\cite{Malgaretti2019}, given a free energy $\beta {F}$ and assuming overdamped Langevin dynamics, one can write the 
{Smoluchowski} equation for the evolution of the probability density function of the stochastic process. From there, one derives a differential equation obeyed by the MPT as well as a cascade of corresponding equations for all higher moments of the (First) Passage Time Distribution. These equations lead to the explicit expressions for the first two moments, $T_1$ (the Mean Passage Time) and $T_2$ (the Passage Time Variance), that we adapt to our specific case.\\At variance with Ref.\cite{Malgaretti2019}, we consider a stochastic process with initial condition $x_0$ and two reflective boundaries at $x_0 \pm L$.  By symmetry the solution has the same form but, as the initial condition is always in the middle of the boundaries, the expressions are considerably simplified. Thus, the Mean Passage Time reads
\begin{equation}
    T_1(x_0) = \frac{1}{D_N} \int \limits_{x_0}^{x_0+L} {dx' e^{\beta {F}(x')}} \int \limits_{x_0}^{x'}{dx'' e^{-\beta {F}(x'')} } 
\end{equation}
while the Passage Time Variance reads
\begin{equation}
    T_2(x_0) = \frac{2}{D_N} \int \limits_{x_0}^{x_0+L} {dx' e^{\beta {F}(x')}} \int \limits_{x_0}^{x'}{dx'' ~T_1(x'') e^{-\beta {F}(x'')} } 
\end{equation}
where $D_N$ is the bulk diffusion coefficient. The coefficient of variation $\gamma$ is defined as
\begin{equation}
    \gamma = \sqrt{\frac{T_2-T_1^2}{T_1^2}} = \frac{\sigma_{T_1}}{{T_1}}
\end{equation}
where $\sigma_{T_1}$ is the {square root of the variance 
of the passage time distribution}. The last equality allows to compute the coefficient of variation from numerical data.

\subsection{Comparison with a previous approach}


We report here briefly a comparison between the results of the current {model and the one} reported in Ref.\cite{bianco2016}.

\begin{figure}[h!]
\centering
 \includegraphics[width=0.4\textwidth]{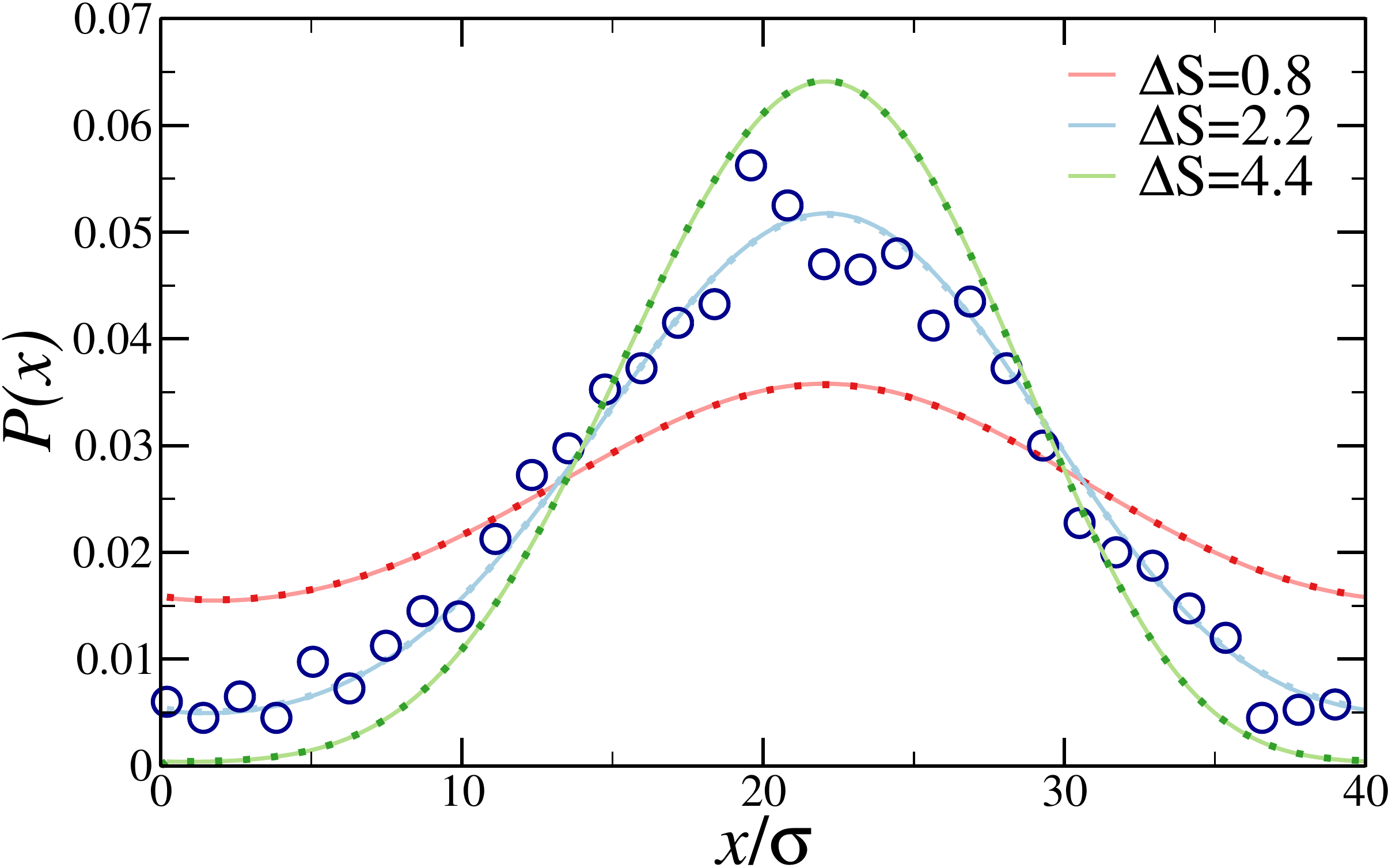}
 \includegraphics[width=0.4\textwidth]{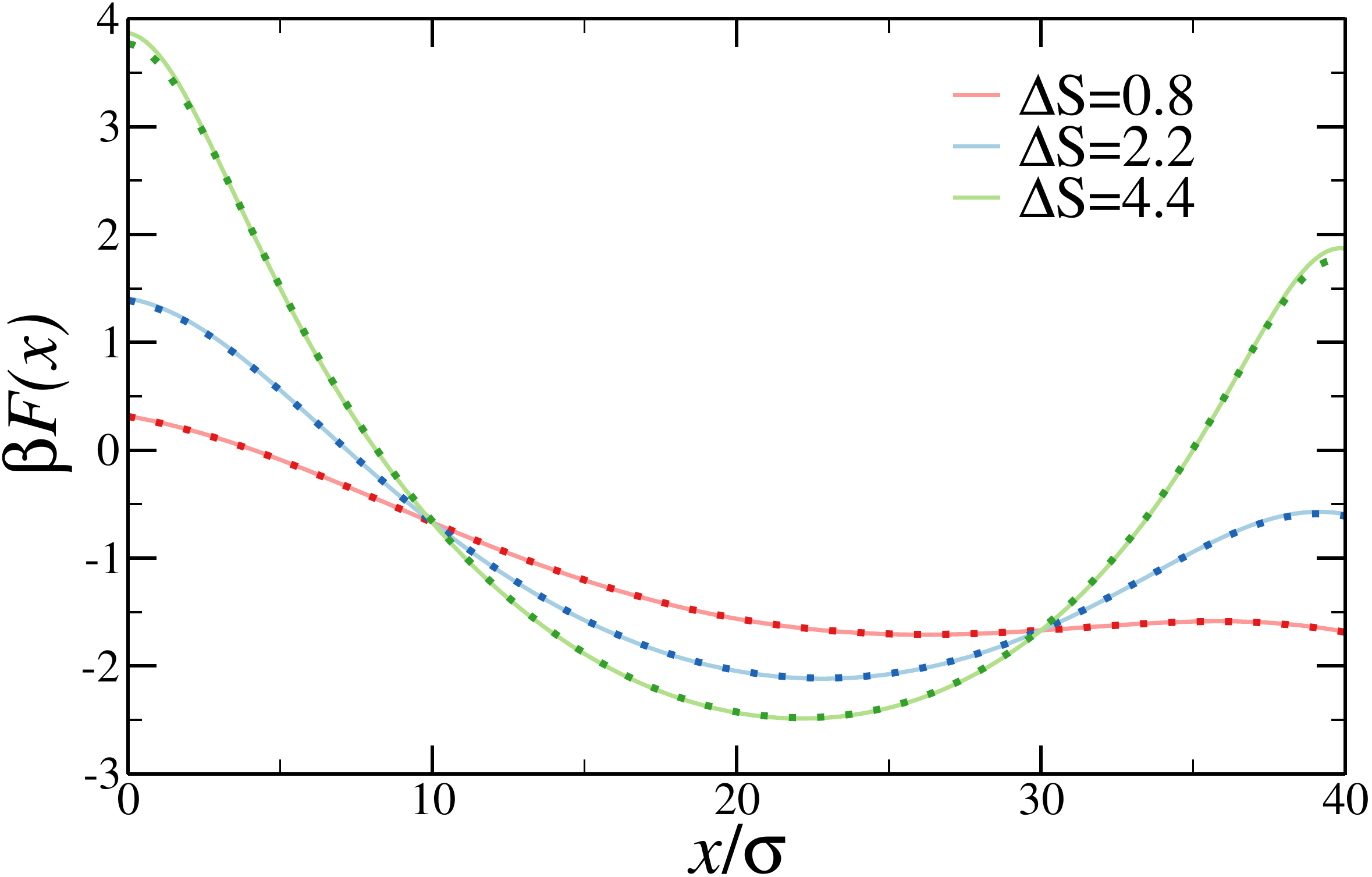}
\caption{(left) Probability distribution of the position of the centre of mass of the chain and (right) Free energy of the polymer as a function of the coordinate $x$, running along the channel axis, for a polymer of length $N=$5 in a channel of length $L=40 \sigma$ and under constant external force $f_0 =0.01 k_B T/\sigma$ and different values of $\Delta S$. Symbols refer to numerical data, solid lines to the approach in Ref.~\cite{bianco2016} and dotted lines to the current approach.}
 \label{fig:comparison_old}
\end{figure}

The research presented in  Ref.~\cite{bianco2016}  primarily focused on driven chains, i.e. in presence of an external force. We choose here to compare the results for very small polymers $N=5$ in presence of a weak external force $f_0 =0.01 k_B T/\sigma$, the same used in the reference work, for a channel of length $L=40\sigma$ and different values of $\Delta S$. We report the comparison in Fig.~\ref{fig:comparison_old}.

In the left panel, we plot the probability of finding the centre of mass at position $x$ and we compare the results from Ref.\cite{bianco2016} (full lines), the current approach (dotted lines) and simulation data (symbols). The old and the current theoretical approach give exactly the same result, that compare very favourably with numerical data. Furthermore, in the right panel, we compare the theoretical free energy: also in this case, the old and the current theoretical approach give exactly the same result.  This comparison further validates our current approach.

\subsection{Scaling of diffusion minima}
\label{sec:diffminima}
We now report  the additional data on the position of the diffusion minima, in particular the first minimum, presenting the data without any rescaling. The main aim  is to demonstrate that the original data sets were very different from each other, thus remarking on the quality of the rescaling.

\begin{figure}[h!]
\centering
\includegraphics[width=0.44\textwidth]{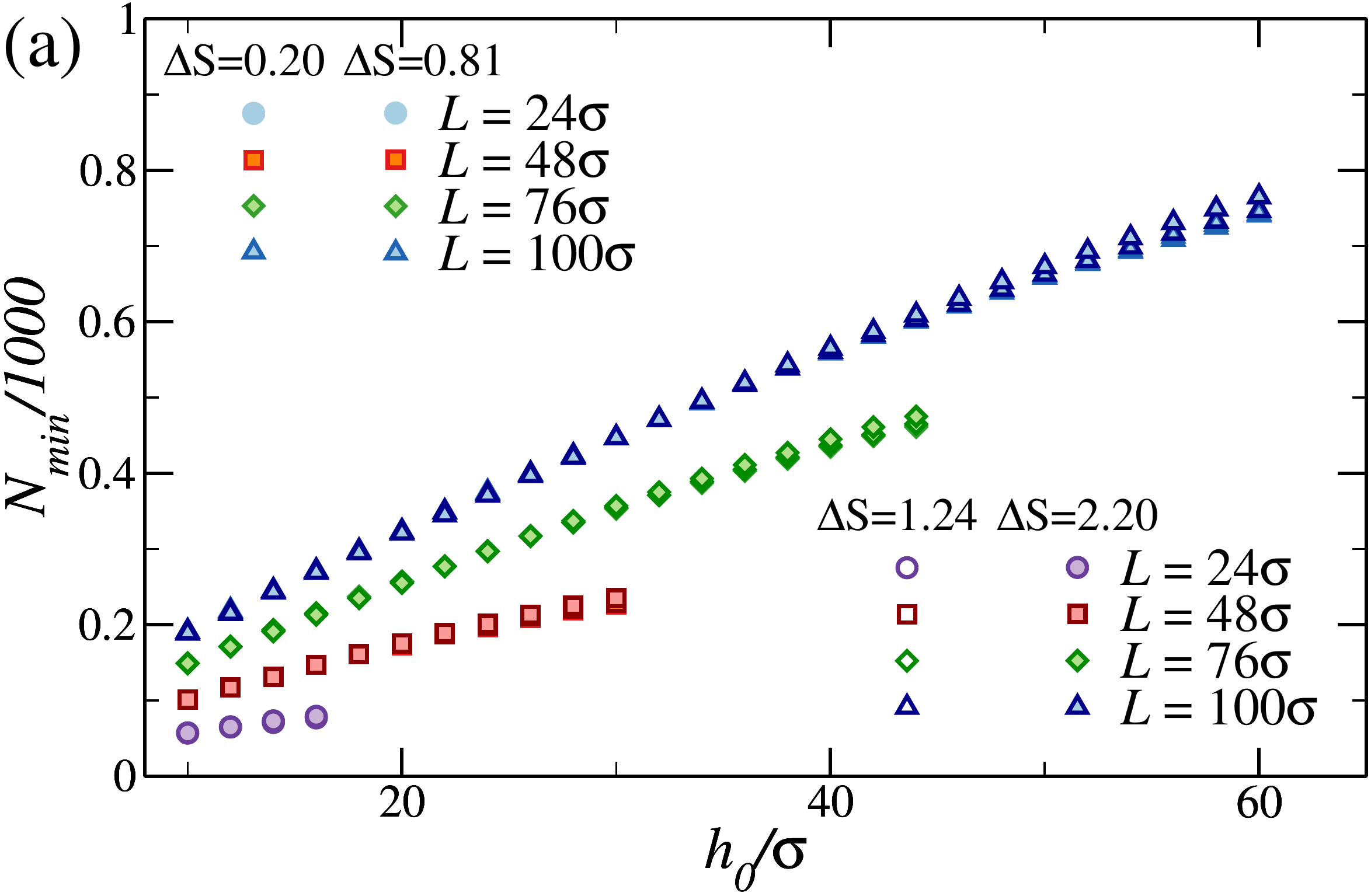}
\includegraphics[width=0.44\textwidth]{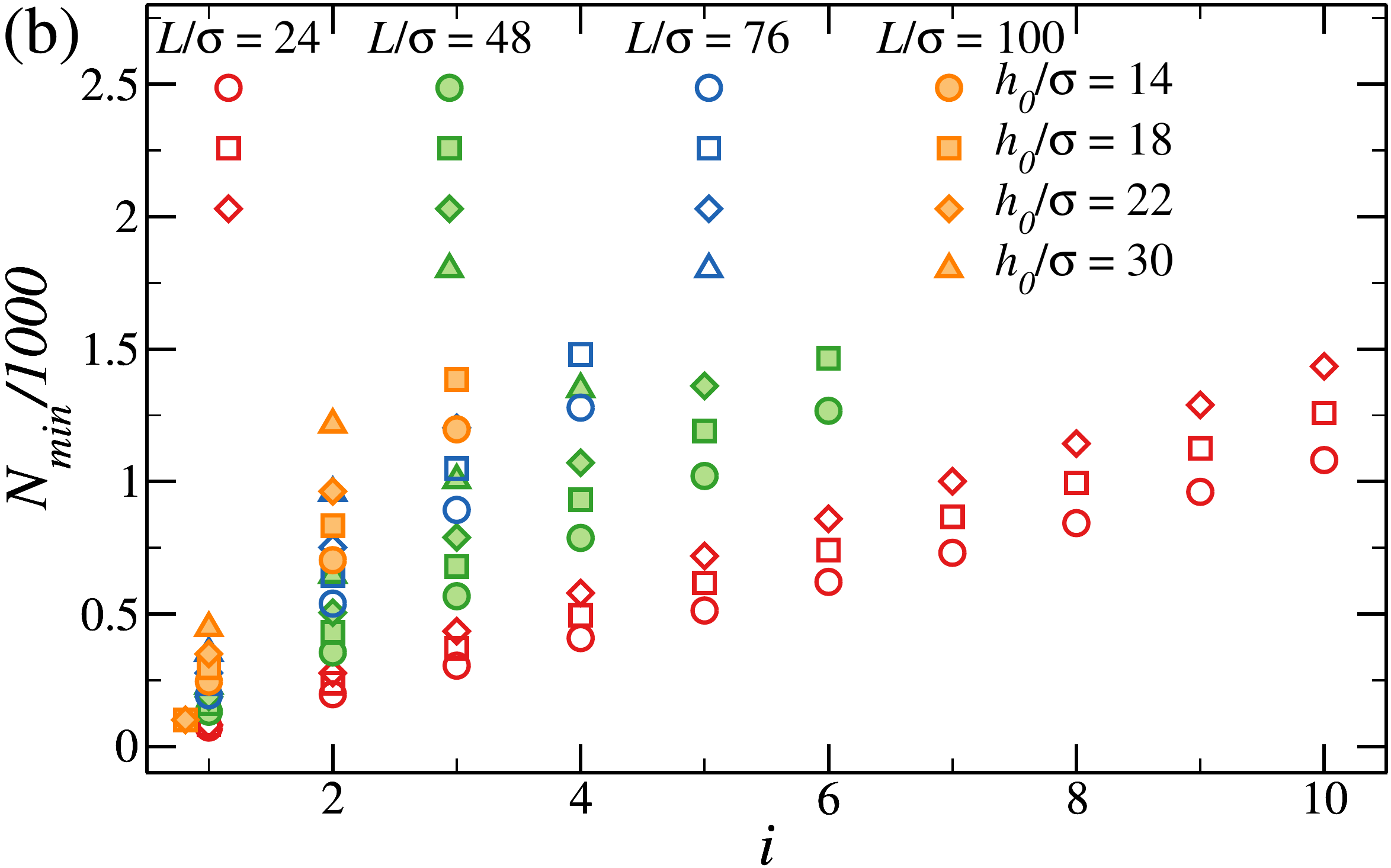}

\caption{(a) Position of the first diffusion minimum as a function of the average channel width $h_0$ for different values of $L$ and $\Delta S$; (b) Position of the diffusion minima as a function of their index $i$ for different values of $h_0$ and $L$ at $\Delta S=$ 0.81. }
 \label{fig:nonscaled}
\end{figure}
In Fig.~\ref{fig:nonscaled}a, we report the position of the first diffusion minimum as a function of the average channel width $h_0$; data reported refer to different values of $L$ and $\Delta S$. We observe that, as mentioned in the main text, the position of the minimum depends strongly on $L$ and $h_0$ and only weakly on $\Delta S$. Next, in Fig.~\ref{fig:nonscaled}b we report the position of the diffusion minima as a function of their index $i$: the reported data  refer to  different values of $h_0$ and $L$ at fixed $\Delta S=$ 0.81. Thus, we  highlight  how the raw data sets depend quite strongly on $L$ and $h_0$, which makes the collapse reported in the main text even more impressive. 

\section{Simulation results and comparison with theory}

\subsection{Simulation details}
\label{sec:sim_model}
We model the polymer as a self-avoiding chain of $N$ monomers in three dimensions, confined in a corrugated channel. We consider self-avoiding polymers, self-avoidance being guaranteed by a repulsive Lennard-Jones interaction, acting between all pairs of monomers
\begin{equation}
 V_{\rm LJ}(r) = 
\begin{cases}
4 \epsilon 
\left[ \left(\frac{\sigma}{r} \right)^{12} -
  \left(\frac{\sigma}{r} \right)^{6} \right] 
+ \epsilon; &{\text {for}}\,\, r<2^{1/6}\,\sigma, \cr
0; &{\text {for}}\,\, r \geq 2^{1/6}\,\sigma,
\end{cases}
 \label{eq:LJ}
\end{equation}
where $\sigma=1$ and $\epsilon=$ 1 $k_BT$, being $k_B$ the Boltzmann's constant and $T$ the absolute temperature. Consecutive monomers along the polymer backbone are held together via a FENE potential 
\begin{equation}
V_{\rm {FENE}}(r) = 
\begin{cases}
- \frac{K}{2} \left( R_{0} \right)^2  \ln \left[ 1 - \left( \frac{r}{R_{0}} \right)^2 \right] ; &{\text {for}}\,\, r \leq R_{0}, \cr
\infty; &{\text {for}}\,\, r > R_{0},
\end{cases}
\label{eq:FENE}
\end{equation}
where $R_0 = 1.5\sigma$, $K=30$ $k_BT/\sigma^2$; such potential prevents the bonds to stretch over the maximum distance $R_0$, hence avoiding crossing events. The confining corrugated channel is modelled as a collection of beads, with diameter $\sigma$, {in contact with each other} and placed around the channel axis according to the following expression
\begin{equation}
 R(x)=\left( \dfrac{ R_{\rm min} + R_{\rm max} }{2} \right) + \left( \dfrac{ R_{\rm min} - R_{\rm max} }{2}\right)\cos\left(\frac{2\pi}{L}x \right),
\end{equation}
where $R_{\rm max}=h_{max}/2$ and $R_{\rm min}=h_{min}/2$ are the largest and smallest radius of the channel, respectively; the main axis of the channel is parallel to the $x$ axis.\\The channel beads are kept frozen during the molecular dynamic simulations in order to guarantee the channel's rigidity. The repulsive potential given by Eq.(\ref{eq:LJ}) applies also between the polymer's monomers and the channel's beads, to guarantee {the impenetrability of the channel walls}. 
The geometry of the channel is captured by a dimensionless parameter, {the \emph{entropic barrier}~\cite{Malgaretti2016}, defined as} 
\begin{equation}
 \Delta S \equiv 2 \log\left[ \frac{R_{\rm max}-\sigma}{R_{\rm min}-\sigma} \right]
 \label{eq:def_deltaS}
\end{equation}
Note that the {maximum} accessible radial distance for any monomer is $R(x)-\sigma$: this justifies {subtracting $\sigma$} in Eq.~(\ref{eq:def_deltaS}). For any channel geometry, we have $ R_{\rm max} + R_{\rm min} = h_0$.\\
We perform standard Langevin Dynamics simulations using the simulation code LAMMPS\cite{LAMMPS}, {integrating the equations of motion with} the Velocity Verlet algorithm (elementary time step of $dt = 10^{-3}$). {We neglect hydrodynamic interactions and investigate the Rouse regime.} We set $\sigma$ as {the unit of length}, $k_BT$ as {the unit of energy} {and the monomer mass $m =$ 1 as the unit of mass}. The unit of time is given by $\tau_0 = 1$; we also set the Langevin friction coefficient $\gamma_L = 1 \tau_0^{-1}$. {We simulate polymers of different {degree of polymerization} $N$, ranging from $N=15$ to $N=200$ and different values of the entropic barrier from $\Delta S=$ 0 ({constant section} channel) to $\Delta S = 1.70$. The {period} of the {{sinusoiodal shape of the}} channel is given by $L$: {we name a single period of the sinusoidal shape a \emph{channel unit.}} Periodic boundary conditions are enforced {along} the $x$ direction: {in order to prevent the self-interactions of the polymer with its own image, induced by the periodic boundary conditions,}  the $x$ size of the simulation box $L_x$ has been adjusted according to the polymer size $N$ fixing $L_x=L$ for $N\sigma\leq L$ (i.e only one channel {unit} is {present} into the simulation box),  $L_x=2L$ for $L<N \sigma\leq 2L$ and so on {(i.e two, or more channels units)}. Statistics are collected, after equilibration, over {$M = 200$ independent runs}, each spanning at least $2\times 10^{9}$ time steps for every value of $N$ and $\Delta S$ considered. More on the length of the simulation runs for a fixed set of parameters can be found in Section~\ref{sec:transloc} }

\subsection{Normalized diffusion coefficient}

\begin{figure}[h!]
\centering
 \includegraphics[width=0.45\textwidth]{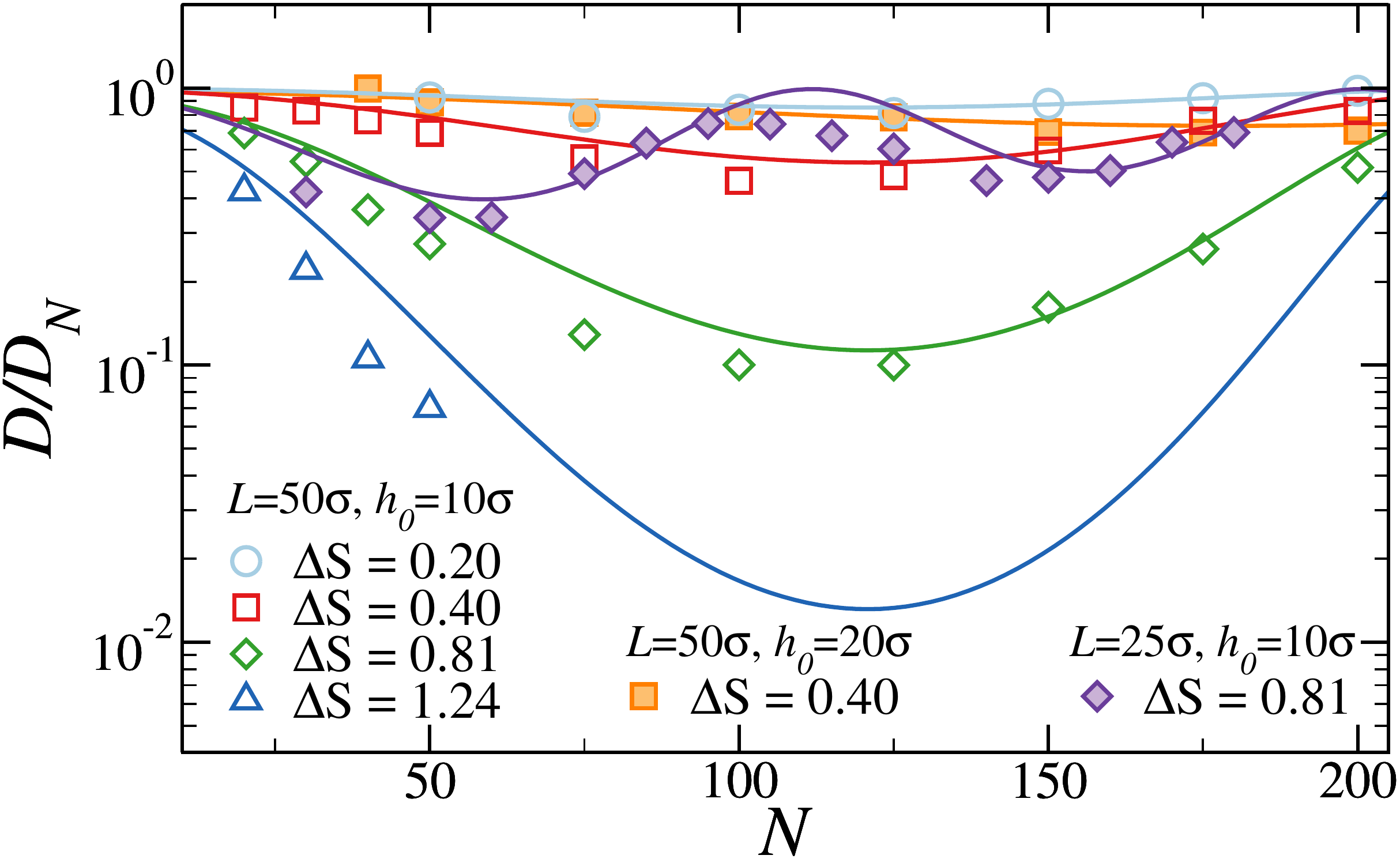}
\caption{Normalized diffusion coefficient $D/D_N$ as a function of $N$ for different values of $L$, $h_0$ and $\Delta S$.}
 \label{fig:normalized}
\end{figure}
We report here briefly the normalized diffusion coefficient $D/D_N$, where $D_N$ is the diffusion coefficient of the whole polymer which, in the Rouse regime, reads $D_N = D_0/N$, $D_0 = k_B T/m \gamma_L$ being the diffusion coefficient of a single monomer. As mentioned in the main text, the non-monotonicity is, in general, more evident for this quantity. 

\subsection{On the magnitude of the end-to-end vector $R_{e}$}
\label{sec:REE}
Similarly to \cite{marenda2017sorting}, we are interested in understanding the extension of the polymer inside the channel, compared to the length scale of the channel corrugation, $L$. As mentioned, in a corrugated channel the magnitude of the end-to-end vector (actually, any observable that depends on the coordinates of the monomers) becomes a function of the position of the centre of mass of the polymer along the channel axis. Nevertheless, following \cite{marenda2017sorting}, we choose here to focus on the average value of $R_e(x)$, $\langle R_e(x) \rangle = R_e$.
\begin{figure}[h!]
  \includegraphics[width=0.45\textwidth]{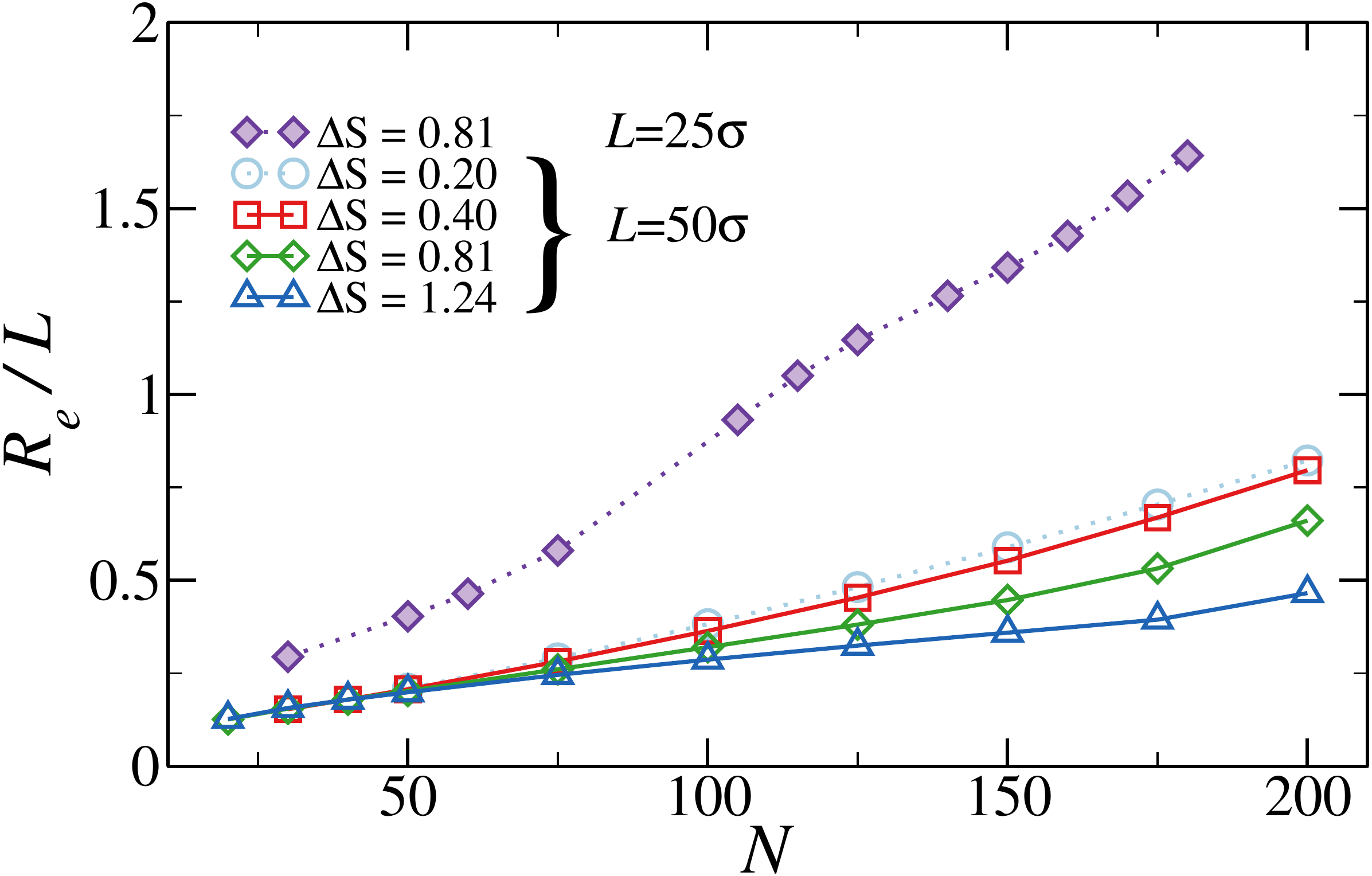}
   \caption{Average root end-to-end mean square distance $R_{e}$ normalized over the length of the modulation $L$ as function of $N$ for different values of $\Delta S$, $L=50\sigma$ and $h_0 = 12\sigma$ or $L=25\sigma$ and $h_0 = 10\sigma$.}
  \label{fig:Reeav}
\end{figure}
We compare the average root end-to-end mean square distance with the length of the channel modulation $L$ for self-avoiding polymers. First, if we compare this data with the results reported in Ref.~\cite{marenda2017sorting} for rings, we notice that linear chains are much more extended than rings: for comparison, $R_e \simeq 50 \sigma$ for a chain of $N=$200 monomers while the longitudinal span, the equivalent quantity for rings, is roughly $25 \sigma$ for a ring of $N=$ 300 monomers. Further, for a channel with corrugation length $L=25\sigma$, we see that we simulate polymers that are on average more extended then $L$. The observations confirm the initial hypothesis behind the theoretical approach: upon increasing $N$, a self-avoiding polymer will increase its size so that it will experience different degrees of confinement at any given position along the channel axis. Further, as reported in the main text, the effective approach is able to maintain predictive power even if the polymer chain is more extended than the channel corrugation (see Fig.\ref{fig:Reeav}). 

\subsection{Mean Passage Times from numerical simulations}
\label{sec:transloc}
We report  the measured values for the mean passage time $T_{1}$, i.e. the average times it takes for a chain to diffuse the length $L$ of an entire corrugation, from the initial position $x_0$ to $x_0 \pm L$. We remark that $x_0$ is chosen at random, after equilibration, thus sampling the equilibrium probability distribution. We will take advantage of this section to briefly discuss a new quantity, $p_{trapped}$, i.e. the fraction of chains that never diffused up to $x_0 \pm L$ within the simulation time, as a key quantity to look at in this kind of systems when assessing if the simulation time was sufficiently long.   

\begin{figure}[h!]
\centering
 \includegraphics[width=0.45\textwidth]{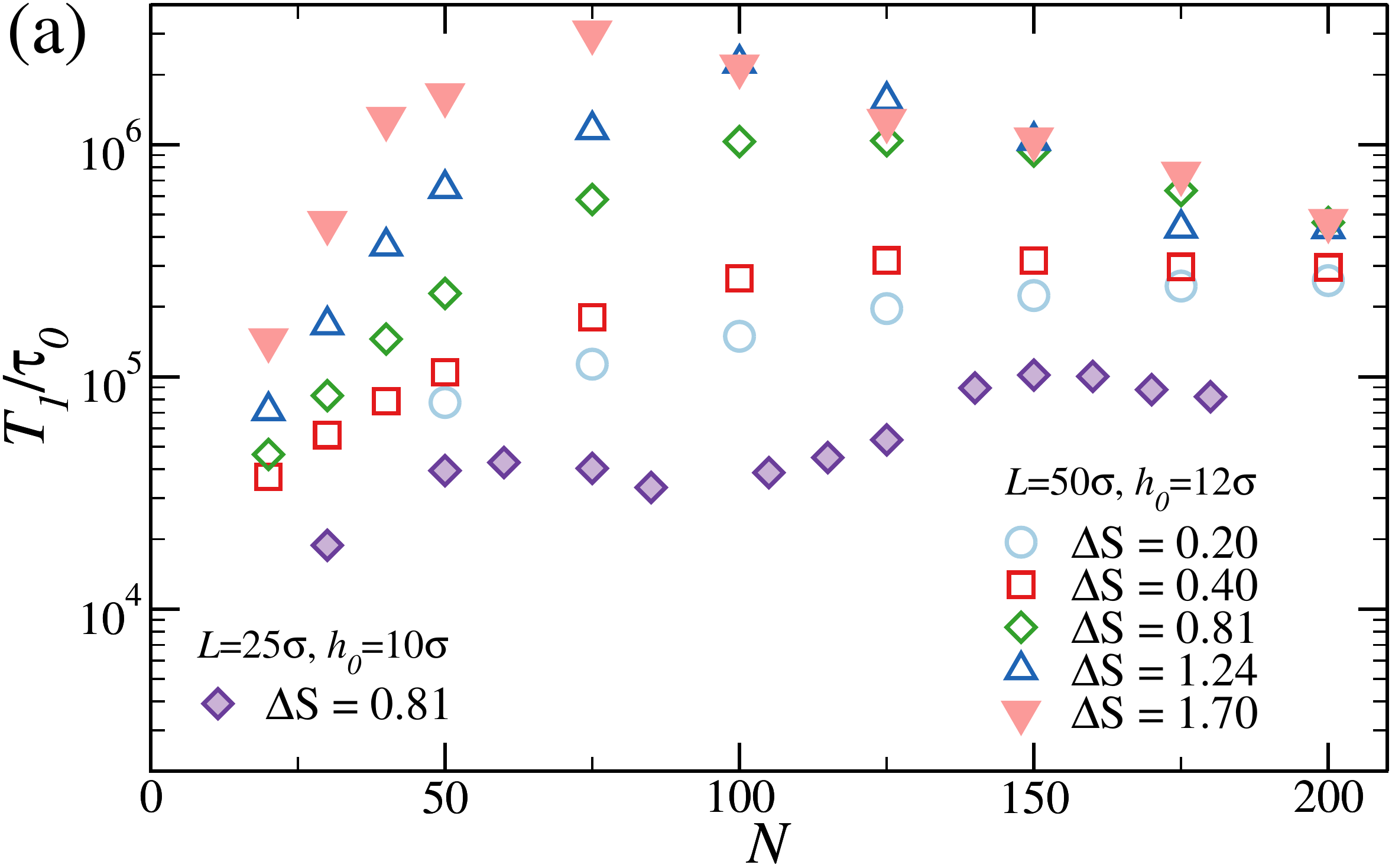}
\includegraphics[width=0.45\textwidth]{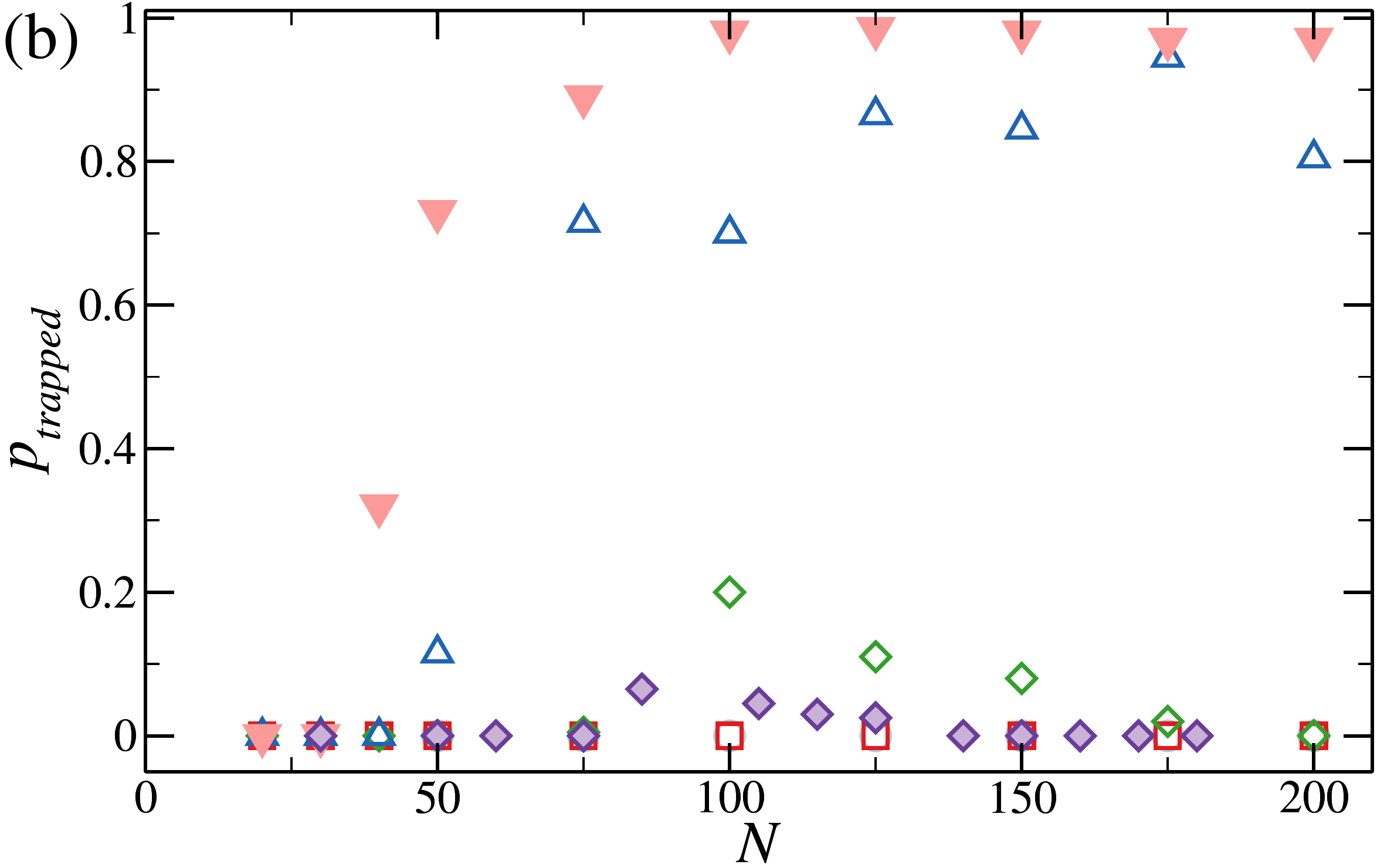}
\caption{(a) Mean Passage Time $T_{MPT}$ (b) Fraction of chains that never diffused up to $x_0 \pm L$ within the simulation time, $p_{trapped}$, as a function of $N$ for different values of $L$, $h_0$ and $\Delta S$. }
 \label{fig:translocations}
\end{figure}

In Fig.~\ref{fig:translocations}, we report $T_{1}$ (panel a) and $p_{trapped}$ (panel b) as a function of $N$. In principle, $T_{1}$ and $D$ essentially hold the same information: indeed, we observe the non-monotonicity in Fig.~\ref{fig:translocations}a. However, we highlight that, for large values of $\Delta S$, the values of $T_1$ seem to become independent on $\Delta S$; further, the peak seems to shift towards smaller values of $N$, in contrast with the theoretical predictions. However, as visible in Fig.~\ref{fig:translocations}b, there is an overwhelming majority (in some cases almost the totality) of chains that never diffused up to $x_0 \pm L$. Thus, in this cases, $T_{1}$ is not a reliable observable, as the estimate is biased by a few outliers. A similar argument can be brought up for the diffusion coefficient: if the majority of the simulated polymers is still trapped within the corrugation, the measured $D$ will not coincide with the real long-time value.\\Indeed, as mentioned in the main text, we take advantage of this measurement for assessing the quality of the numerical diffusion coefficient. For simplicity, we choose to discard the data that does not fit the criterion chosen because, as visible in Fig.~\ref{fig:translocations}b, our data show roughly a bimodal scenario, where either all (200) simulated polymers fitted the criterion or the vast majority of them were trapped inside the initial corrugation for the whole trajectory. We stress that we ran some of the simulation above for a very long time, $10^6-10^8 \tau_{MD}$ which correspond to $10^9-10^{11}$ time steps and a several weeks/months of computational time. Worse, even after this long time, we lacked a reliable estimate of the simulation time needed to reach the true long-time regime. This shows how tremendously expensive these simulations can be, when $\Delta S$ becomes large, which also highlights the usefulness of the proposed theoretical approach.

\providecommand{\noopsort}[1]{}\providecommand{\singleletter}[1]{#1}%

\end{document}